\documentclass{emulateapj}
\accepted{February 13, 2017}

\shortauthors{Kesseli et al.}
\shorttitle{BOSS Empirical Templates}

\begin{document}

\title{An empirical template library of stellar spectra for a wide range of spectral classes, luminosity classes, and metallicities using SDSS BOSS spectra}

\author{Aurora Y. Kesseli\altaffilmark{1}, Andrew A. West\altaffilmark{1}, Mark Veyette\altaffilmark{1}, Brandon Harrison\altaffilmark{1}, Dan Feldman\altaffilmark{1}, John J. Bochanski\altaffilmark{2}}

\altaffiltext{1}{Boston University Astronomy Department, 725 Commonwealth Ave, Boston, MA 02215}
\altaffiltext{2}{Rider University, 2083 Lawrenceville Rd, Lawrence Township, NJ 08648}

\email{aurorak@bu.edu}

\begin{abstract}

We present a library of empirical stellar spectra created using spectra from the Sloan Digital Sky Survey's Baryon Oscillation Spectroscopic Survey (BOSS). The templates cover spectral types O5 through L3, are binned by metallicity from -2.0 dex through +1.0 dex and are separated into main sequence (dwarf) stars and giant stars. With recently developed M dwarf metallicity indicators, we are able to extend the metallicity bins down through the spectral subtype M8, making this the first empirical library with this degree of temperature \emph{and} metallicity coverage. The wavelength coverage for the templates is from 3650 \AA\ through 10200 \AA\ at a resolution better than R$\sim2000$. 
Using the templates, we identify trends in color space with metallicity and surface gravity, which will be useful for analyzing large data sets from upcoming missions like LSST. 
Along with the templates, we are releasing a code for automatically (and/or visually) identifying the spectral type and metallicity of a star. 

\end{abstract}

\maketitle

\section{Introduction} 

Empirical stellar spectral libraries are crucial for many areas of astronomical research that include, but are not limited to, simple assignment of stellar spectral types by-eye, more complicated machine learning spectral typing, and modeling the spectral energy distributions and stellar populations of galaxies.  
The determination of stellar properties such as surface gravity, metallicity and effective temperature is often completed by comparisons to empirical templates with known parameters, or modeled spectra. 
Stellar template libraries are also an important teaching resource, from examples of stellar spectra for introductory classes, to detailed radiative transfer at the graduate level.

With the advent of the Large Synoptic Survey Telescope \citep[LSST; ][]{ivezic08} and many other large surveys, photometric data sets of unprecedented size will be available to astronomers. LSST will not include a spectrograph, and so the characterization of stellar parameters must be based entirely on the available photometric data. 
In preparation for LSST, \citet{miller15} created a machine learning technique, which was taught to determine metallicities solely from photometric colors, using Sloan Digital Sky Survey (SDSS) stellar spectra with log $g$, [Fe/H], and $T_{eff}$ measurements from the Segue Stellar Parameters Pipeline \citep[SSPP; ][]{Lee08}. The pipeline, however, is only quoted to be accurate for spectral types F through mid-K (outside this temperature range may be accurate but requires further testing), excluding a large portion of stellar parameter space. 
In the modern age of these large surveys, data sets will be so large that individual analyses will be infeasible, and machine learning or other statistical techniques will become increasingly important. To accurately train machines, datasets with known properties that cover the entire range of parameter space is needed.

Empirical spectral libraries have a rich history in stellar astronomy (e.g., An Atlas of Digital Spectra of Cool Stars: \citealt{turnshek85}; Pickles: \citealt{pickles08}; ELODIE: \citealt{pru01}), but each has limitations, and none of them include a range of M star properties. Because of the complex molecular rich atmospheres of low-temperature stars, estimates of log $g$ and [Fe/H] have only become accurate within the last few years \citep[e.g.,][]{rojas10, terrien12, mann13, newton14}, and were therefore unavailable when these libraries were assembled. 
The ELODIE library is also limited by its small wavelength coverage (3900 -- 6800 \AA), which does not include many of the longer-wavelength features necessary for studying low-mass stars' metallicities and surface gravities (e.g., Na I at 8200 \AA). 
Numerous groups have used SDSS to create partial libraries of specific effective temperature regions because of the wide wavelength coverage and vast number of spectroscopic and photometric observations available. In the solar mass regime (4,500 -- 7,500 K), \citet{Lee08} created the Sloan Extension for Galactic Exploration and Understanding (SEGUE) SSPP and estimated $T_{eff}$, log $g$, and [Fe/H] for over 100,000 stars. In the low-mass regime, 
\citet{hawley02} first assembled a sample of M, L and T dwarf single object spectra. \citet{bochanski07} improved upon the sample by creating co-added templates for active and non-active low-mass stars (M0-L0), and \citet{schmidt14} extended the templates through the spectral subtype of L6. \citet{sav14} expanded the library further by compiling co-added templates of low-mass (and low-metallicity) subdwarfs. However, there is no single source library that covers parameter space in the low-mass end (M dwarf metallicities and surface gravities) through the high-mass regime. 

Synthetic template libraries are also used widely throughout astronomy and are extremely useful tools, since the parameters like log $g$, [Fe/H], and $T_{eff}$ are model inputs and can span a wider parameter space than any empirical template library. Many of these complete model libraries are available to the public (e.g., Kurucz: \citealt{kur79}; BT Settl: \citealt{allard12}). However, the models must use simplifications such as plane parallel, local thermal equilibrium, etc., to create these synthetic spectra \citep{husser13}. Synthetic stellar spectra are also limited because of incomplete lists of line opacities, and a lack of knowledge of input parameters for pressure broadening models \citep{allard12, husser13}. This is especially problematic for the complex, molecular-rich atmospheres of low-mass stars, where the model spectra do not exactly match the observed spectra across all wavelengths. 
Empirical libraries are therefore still extremely important both to constrain models and to use in regimes where the models are limited. 

With a new, complete stellar template library, spectral typing and determinations of other important stellar properties will improve. In addition, the templates can be used in more complicated stellar population modeling. 
Stellar population synthesis models have long been used as a tool to assemble model galaxies from the co-addition of  individual stellar spectra \citep[e.g., ][]{bruz83, bruz03}. Many studies have used these model galaxies to determine basic galactic properties such as mass-to-light ratios (M/L) and to refine the Tully-Fisher relation \citep[e.g., ][]{bell01}. 
Recently, this method has been used to study the initial mass function (IMF) in distant galaxies and has led to a growing body of evidence that the IMF is not constant as previously thought, but can change depending on the environment \citep[e.g., ][]{van11, conroy12, geha13}. The debate over the form of the IMF is concentrated in the low-mass end of the stellar sequence ($<1 M_{\odot}$), where estimates of its form produce different numbers of low-mass stars that span at least an order of magnitude. 
To disentangle the low-mass stars from the spectrum of an elliptical galaxy, whose light is completely dominated by red giants, \citet{van11} examined absorption features (Na I and FeH) that are prominent in low-mass, main-sequence stars and inferred the abundance of low-mass stars, and hence an IMF. A leading hypothesis for the physical parameter that best correlates with changes in the IMF is metallicity.  To explore this scenario further, we need a template library with high quality low-mass stars that include metallicity variations and a wavelength coverage that includes these dwarf sensitive features (Na I at 8200 \AA\ and FeH at 9919 \AA).




In this paper, we present a library of empirical stellar spectra for spectral types O5-L3. Our library separates out luminosity classes (dwarf and giant) for spectral types A0 through M8, and contains metallicity bins for spectral types A3 through M8. We do not include white dwarfs in our sample; several catalogs of SDSS white dwarfs can be found in other studies \citep[e.g., ][]{klein13}. The templates were created by the co-addition of individual stellar spectra from SDSS's Baryon Oscillation Spectroscopic Survey (BOSS). BOSS contains over 500,000 well-calibrated stellar spectra spanning wavelengths 3,600 -- 10,400 \AA\ at a resolution of $R \sim 2000$ \citep{dawson13}. This stellar template library spans stellar parameter space and contains wavelength coverage that is not available from other empirical libraries. In addition, because each object has photometry associated with it, we can use this library to identify empirical photometric relations among spectral type, metallicity, surface gravity and color. Along with the templates, we are also releasing a revised version the ``Hammer" spectral type facility \citep{covey07}, dubbed ``PyHammer". The original code assigns an automatic spectral type by measuring a variety of spectral lines/features and performing a least-squares minimization. The code also allows the user to complete visual spectral fitting. We have rewritten the code in Python using our templates for comparison, and augmented the code to assign a metallicity, in addition to a spectral type. We also include a GUI for visual spectral typing.

In the following sections, we present the methods for determining the spectral type, the radial velocity (Section \ref{RV}), the surface gravity (Section \ref{logg}), and the metallicity (Section \ref{metallicity}). We then describe the co-addition process (Section \ref{coadd}). Next, we describe the results from the spectroscopic data of the templates (Section \ref{Results:spec}), and the photometric data (Section \ref{Results:phot}). Finally we draw conclusions for our templates in Section \ref{conclusions}. We also include an appendix, which describes the ``PyHammer" spectral typing code released along with this paper.

\section{Methods} 

We selected and co-added spectra in bins of metallicity ([Fe/H]), surface gravity and spectral type to create our empirical catalog. Since no universal pipeline or technique exists for determining metallicities and surface gravities over the full range of stellar temperatures, we used a combination of techniques to extract this information from individual spectra, dependent on their spectral type. We separated the spectra into the following six spectral classification categories, each of which have a different method for the determination of parameters: O and B stars; A stars; F, G and early K stars; late K and early M stars; and late M stars. Our catalog contains templates for early-type L stars as well, but we did not separate by surface gravity or metallicity for the L subclasses because the majority of the metal sensitive and surface gravity sensitive features are in the infrared. \citet{schmidt15} assembled a library of ultracool dwarfs, which uses both original SDSS spectra and newer BOSS spectra, and has been spectroscopically classified (using the original ``Hammer" program). We therefore selected all of the spectra taken with BOSS, and used those for our L dwarf templates.

To ensure the quality of all of the spectra and photometry, we applied some basic quality cuts within our initial SDSS CasJobs Query\footnote[1]{http://skyserver.sdss.org/CasJobs}. 
We measured the signal-to-noise in each of the five photometric bands. We included the object if the median value of the five signal-to-noise values (from each band) was greater than ten. 
We also required that the spectra to be classified as a star (not a galaxy or QSO), and to be taken with the BOSS spectrograph. We required that the errors in the photometry be less than 0.1 mag in each individual band that is included in the final co-added photometry, and that the photometry in that band was not flagged for being deblended, containing a cosmic ray, or saturated. Finally, we required that the extinction in the \textit{r}-band be less than 0.25 mag for the object to be included, which we found effectively removed any spectra that were visibly altered by extinction.

Once the spectra were selected, we visually classified each of the spectra using the ``Hammer" spectral type facility \citep{covey07}. We then calculated a radial velocity (Section \ref{RV}) to shift each spectrum into its rest frame before further spectral analysis and/or co-addition. We separated the spectra into metallicity (Section \ref{metallicity}) and surface gravity (Section \ref{logg}) bins when applicable. 
There are some holes in our template coverage across parameter space, where we could not find BOSS spectra at all the metallicity/spectral type/surface gravity values.
Once segregated, each spectrum was cataloged and co-added to create empirical spectral templates (Section \ref{coadd}).


\subsection{Radial Velocity}
\label{RV}

\subsubsection{O, B and A Stars} 
\label{OBA:RV}

For the O, B, and A stars, we used the spectroscopic radial velocities from the SDSS spectroscopic reduction pipeline \citep{stoughton02}. 
The \textsc{spectro1d} pipeline performs two estimates of radial velocities: a cross-correlation with stellar templates from SDSS commissioning spectra with known radial velocities, and a comparison of emission lines to common galaxy and quasar emission lines. The final radial velocity is chosen from the method with the highest confidence level. 
The quoted accuracy is about 10--15 km $s^{-1}$ \citep{adelman08}. 

\subsubsection{F, G, and Early-Type K Stars} 
\label{FGK:RV}

Stars with effective temperatures between 4,500 -- 7,500K ($\sim$K4-F0) fall within the temperature range where \citet{Lee08} quote that the SSPP produces reliable stellar parameters (e.g., metallicities, surface gravities, effective temperatures, and radial velocities). We used the radial velocities determined by the SSPP for these intermediate temperature stars. The SSPP determined the radial velocity shifts by cross correlating the templates with the ELODIE templates \citep{pru01}, a method which was shown to be more reliable than the original SDSS spectroscopic reduction pipeline in \citet{adelman08}. However, the ELODIE library is not complete, and if no matching template existed, the SSPP adopted the original radial velocity determined by the \textsc{spectro1d} pipeline.  The radial velocities determined by cross correlation to the ELODIE templates have reported uncertainties of 5--9 km $s^{-1}$ \citep{adelman08}. 

\subsubsection{Late-Type K, M, and L Stars} 
\label{KM:RV}

To determine the radial velocities for the lowest-temperature stars, we chose to perform our own cross correlation with the M dwarf template library from \citet{bochanski07}, for spectral subclasses K5 through M9, and \citet{schmidt14} for the L dwarfs. The cross correlation method examines three regions of the spectrum (5000--6000 \AA, 6000--7000 \AA, and 8000--9000 \AA), and for each region performs a cross correlation. The minimum of the cross correlation function is recorded, and a sigma clipped median of the measurements from all the regions is reported as the radial velocity. We calculated an uncertainty from the standard deviation of all the individual radial velocity measurements (in each region) added in quadrature with the precision at which radial velocities can be calculated from BOSS's resolution. We take this radial velocity precision to be $\sim 7$ km $s^{-1}$ as determined by \citet{bochanski07} in a comparison with high precision radial velocity measurements of Hyades cluster members. In the comparison with the Hyades cluster, 
\citet{bochanski07} found radial velocities that were on average 4 km $s^{-1}$ more precise than the standard SDSS pipeline's radial velocities.
For the templates with a signal-to-noise ratio (SNR) $>10$ we find a median uncertainty of $9 \pm 0.5$  km $s^{-1}$.  The radial velocity code used to compute these values will be available with the PyHammer code. Further descriptions of the code, and tests of its accuracy are given in the Appendix.

\subsection{Surface Gravity} 
\label{logg}

\subsubsection{O and B-type stars}
For the highest temperature stars (O and B), we did not separate by surface gravity. 
Even though we do not distinguish between dwarf and giant stars, we can eliminate white dwarf contaminants from our sample, which are the vast majority of non-main-sequence stars in this temperature regime \citep{bressan12}. To address this contamination, we cross matched our sample with the \citet{klein13} catalog of SDSS white dwarfs to remove the majority of the contaminants. We also spectroscopically vetted our sample of white dwarfs during the visual spectral classification process, removing any sources with significant line broadening associated with the extreme surface gravities of white dwarfs. This white dwarf vetting process was also applied to A-type stars.

\subsubsection{A-type Stars}

According to stellar evolution models, giants do not spend much time in the A star color regime \citep{bressan12}. In a uniform sample, we could assume that the majority of the A stars in our sample were main sequence stars and would not need to be separated by surface gravity. However, due to the SDSS target selection, previous studies have shown that 
a significant portion of halo stars with A star colors are giants called blue horizontal branch stars \citep[BHB; ][]{morrison00,helmi03, santucci15}, making surface gravity estimates necessary. Because A stars do not fall within the conservative temperature range for which \citet{Lee08} state that the SSPP parameters are accurate, \citet{santucci15} independently determined the surface gravities for over 10,000 stars with A-star colors in SDSS by measuring the width of the wings of the Balmer absorption lines. Stars with higher surface gravities should exhibit noticeably broader lines because of increased pressure broadening. \citet{santucci15} then compared the results of their surface gravity values to the results reported by the SSPP and found $92-95\%$ of stars fell into the same category (dwarf or giant), thus extending the temperature range for which the SSPP can be trusted to distinguish main-sequence from giant A-type stars. We therefore adopted the SSPP method and separated our A stars into a main sequence and a giant bin, using the same criteria listed in \citet{santucci15}, where a BHB giant had log $g < 3.8$ and a main sequence star had log $g > 3.8$. For more information about how the SSPP calculated surface gravities see Section \ref{FGKlogg}.

\begin{figure}[]
\begin{center}
\includegraphics[scale=0.44]{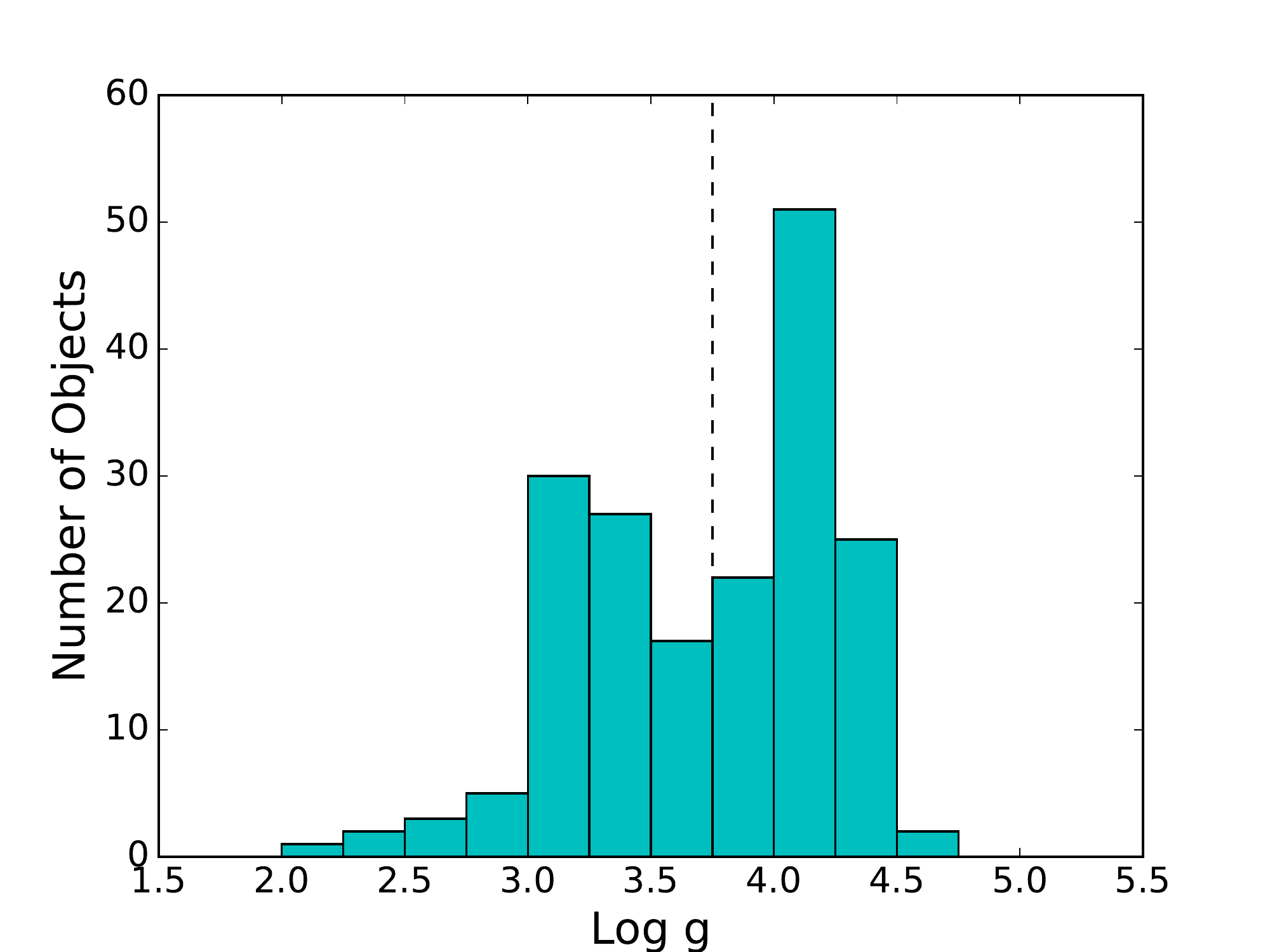}
\caption{\small
Log $g$ distribution of all the stars selected (and spectral typed) as A stars. There are two distinct populations: one centered around log $g = 3.2$ and one around log $g = 4.0$. The population with lower surface gravity is consistent with blue horizontal branch giants, while the higher surface gravity population is composed of main sequence stars and/or blue stragglers. The dashed line at log $g = 3.8$ marks the distinction between main sequence stars (to the right of the line) and BHB giants (to the left of the line). 
}
\label{f:Alogg}
\end{center}
\end{figure}

In Figure \ref{f:Alogg}, which shows the distribution of A star surface gravities, we see two distinct populations of A-type stars. The distribution is very similar to the previous work on surface gravities of A stars by \citet{santucci15}. Figure \ref{f:Alogg} also validates our visual vetting process to remove white dwarfs, since there are no stars with surface gravities in the white dwarf range (typically log $g > 7.0$) in our sample. We can then be relatively confident that white dwarf contamination in our O and B stars is also minimal. 

\subsubsection{F, G, Early-Type K Stars} 
\label{FGKlogg}

The SSPP computes an extremely robust calculation of stellar parameters based on both SDSS spectroscopy and photometry. 
Reported surface gravities from the SSPP use ten different methods, each of which determines a surface gravity estimate.  Most of these methods are well tested, and include: minimization of spectra and photometry compared to models \citep{kurucz93, allende06}, neural network approaches \citep{refi07}, spectral line fits as a function of broadband colors \citep{wilhelm99}, and measurements of well known surface gravity sensitive lines \citep[Ca I: 4226.7\AA, MgH: 5200\AA; ][]{morrison03}. After a surface gravity is returned by each of these methods, statistical outliers are removed from the sample and the final bi-weight average of the log $g$ of the star is returned \citep{Lee08}. 
The SSPP returns surface gravities to an accuracy of $\sigma$(log $g) = 0.21$ dex within the spectral range of F, G, and early K stars \citep{Lee082}. 

We split the spectra into two different surface gravity bins representative of giants and dwarfs. The Padova stellar evolution tracks \citep{bressan12} specify a temperature, surface gravity, and evolutionary phase (e.g., subgiant, giant, supergiant, white dwarf) for each initial mass. Using this information, we chose to call anything with $6.0 >$ log $g > 3.8$ a main sequence dwarf star, and anything with log $g < 3.2$ a giant star. We do not include any stars with log $g$ values between 3.2 and 3.8, but keep this range as a buffer zone, because stars with different temperatures turn off the main sequence at slightly different log $g$ values. Anything with log $g > 6.0$ is considered a white dwarf. We did not separate into giants, subgiants and supergiants because any further separation resulted in too few spectra for each bin.

\subsubsection{Late-Type K, All M Stars} 
\label{Mlogg}

To distinguish low-temperature dwarfs from giants we followed the steps in \citet{bochanski14}. Since giants are intrinsically bright, and thus have to be at large distances in SDSS, they will not exhibit observable proper motions. To isolate main sequence stars, we required that the proper motion in RA or Dec be greater than two times the error in the proper motion, while the giant stars must have proper motions less than two times the error in the proper motion. This proper motion cut however, introduces a high percentage of extra-galactic source contamination in the giant category because elliptical galaxies' spectra often resemble M stars (since M giants are the main observable constituent), and also will not have any measurable proper motion. \citet{bochanski14} note this problem in isolating M giants, and use a $g-i, i-K$ color cut \citep[Equation 5 in ][]{bochanski14}, which \citet{peth11} show effectively separates the two populations. Main sequence and giant stars also occupy distinct locations on the $J-K, J-H$ color diagram \citep{bessell88}. To use the IR color cuts \citep[Equations 1 -- 3 in][]{bochanski14} that make use of the dwarf-giant distinction in IR color space, we matched all of our SDSS targets with the corresponding targets from the Two micron All-Sky Survey (2MASS), which provided us with $J$, $H$, and $K$-band photometry for each object. \citet{bochanski14} conclude that using all of the M giant photometric and proper motion cuts returns approximately $\sim 20 - 50 \%$ M giants (contaminants include: dwarfs, galaxies, other sources). Contamination in the M dwarf category should be minimal, since any giant close enough to have a measured proper motion would saturate the SDSS detectors. 

To remove the remainder of the non-giant contaminants from our M giant category, we resorted to visual inspection of the spectra. We found a few carbon star and extragalactic sources, which could easily be identified spectroscopically. To spectroscopically distinguish dwarfs from giant stars, we focused on the gravity sensitive features, Na I (8200 \AA) and the calcium triplet ($\sim$8600 \AA). The Na I doublet is extremely sensitive to gravity \citep{schlieder12} and is only strong in dwarf stars, while the calcium triplet is more prominent in giant stars \citep{jones84}. After our spectroscopic classification, we conservatively removed $\sim75\%$ of the selected stars and we expect the vast majority of our remaining spectra are true M giants. 

\subsection{Metallicity} 
\label{metallicity} 

\subsubsection{O and B-type stars}
We do not separate O and B-type stars by metallicity. The metallicity history of the disk of the Milky Way is well constrained in time \citep{hayden15}, with recent measurements of metal content being roughly solar metallicity. Since O and B stars have such short lifetimes ($\sim5-500$ million years) and the majority of stars are formed within the disk, we can assume that the majority have metallicities ([Fe/H]) between 0.0 and 0.5 dex. 

\subsubsection{A-type Stars}
\label{A:metal}

\begin{figure}[]
\begin{center}
\includegraphics[scale=0.47]{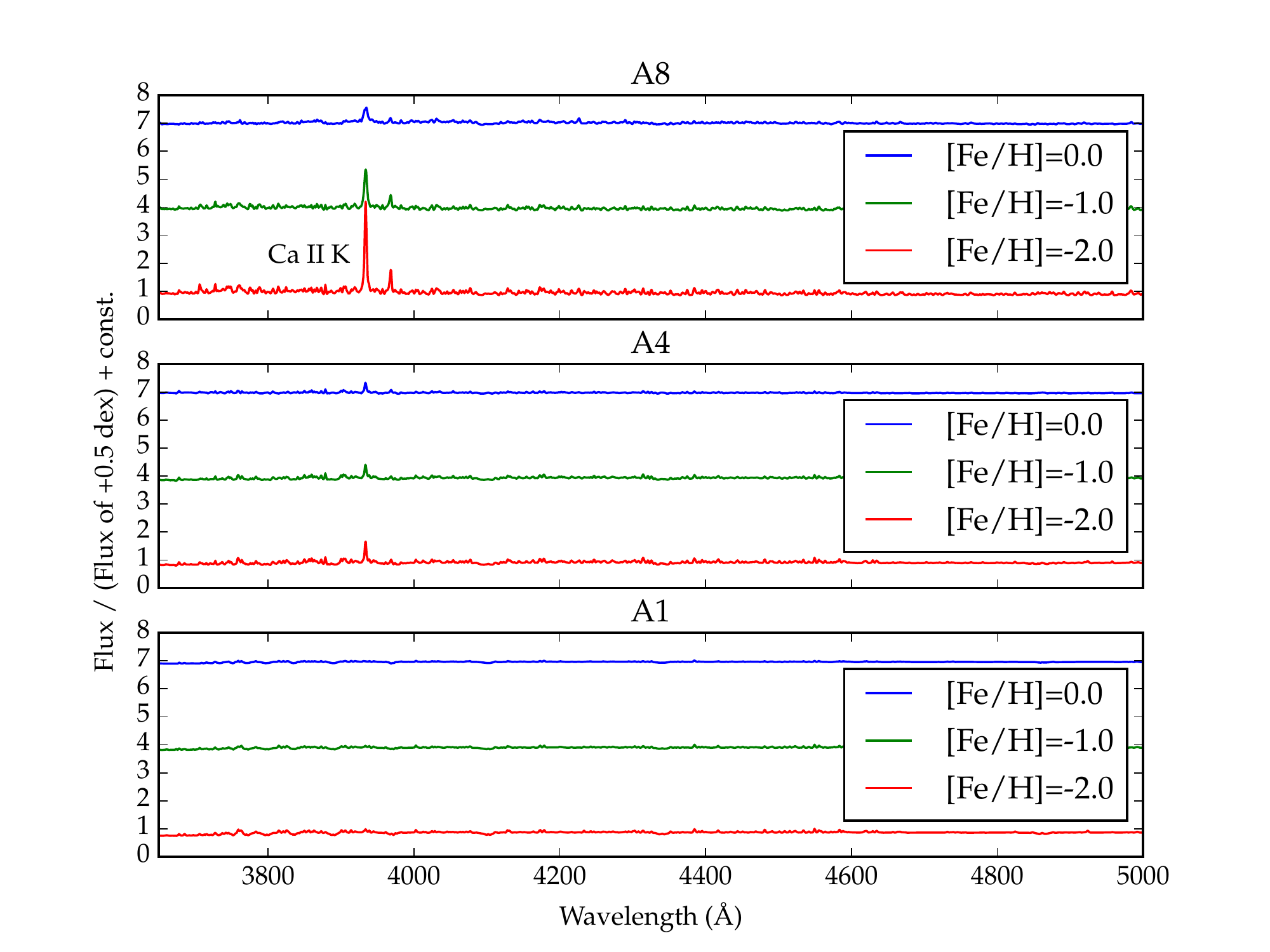}
\caption{\small
Model stellar spectra of different metallicities divided by the corresponding model spectrum at [Fe/H] = 0.5 for different spectral subtypes \citep{allard12}. The Ca II K line shows the largest dependence on metallicity for A stars. For the A8 (top) and A4 (middle) stars, there are significant changes between the metallicity bins. While the A1 (bottom) spectrum does show changes in Ca II K line depth, they are not significant enough changes for us to confidently separate out metallicity bins based on the equivalent width of the Ca II K line.
}
\label{f:Adivision}
\end{center}
\end{figure}

Because A stars have significant lifetimes (up to a couple billion years), we expect there to be a larger spread in metallicities than found in O and B stars. Previous studies of A stars within SDSS have also shown a much larger spread in metallicity than the metallicity histories of the Milky Way disk predict \citep[e.g., ][]{helmi03}, making it important to determine metallicity estimates for these stars. Since the halo does not have ongoing star formation, we would expect the only low-metallicity stars to be low-mass stars, because A stars formed when the Galaxy was on average lower-metallicity would have evolved off the main sequence. However, low-metallicity ([Fe/H] $\sim-2.0$) A stars with surface gravities comparable to main sequence stars (dubbed ``blue stragglers") are observed in the halo with high frequencies \citep{helmi03, santucci15}. 
Because A stars have effective temperatures $> 7500$ K, and are therefore outside the quoted accuracy limits of the SSPP, we tested the SSPP metallicity outputs by comparing to synthetic stellar models. We used the Phoenix models \citep[BT Settl grid; ][]{allard12} and compared different metallicities of the same temperature to determine the most metal sensitive lines. Synthetic models of A stars are reliably accurate because their absorption profiles are relatively simple (do not contain molecules or dust), yet they are not quite hot enough to be dominated by non-local thermodynamic equilibrium metal line-blanketing \citep{auf01}. Figure \ref{f:Adivision} shows the Ca II K line (3933.7 \AA), which is by far the most metal sensitive line and remains metal sensitive for temperatures from 7400 K -- 9800 K. Above 9800 K, the effective temperature is so high that the Ca II K line is weak and the depth of the Ca II K absorption line is on the same order as the uncertainties, making metallicity estimates for A0 through A2 stars unreliable. 

We created metallicity estimates by determining the equivalent width of the Ca II K line for different metallicity BT Settl models \citep{allard12}. We used models with a log $g$ value of $4.0$. To determine the equivalent widths of the BT Settl models, we first changed the resolution to match that of BOSS (R=2000). We processed all of the models through the ``Hammer" spectral typing facility to accurately compare the effective temperature based models to the the spectral typed BOSS spectra. Effective temperature and spectral type should be roughly one to one, but are not necessarily regularly spaced throughout the entire stellar spectral sequence. 
To construct the metallicity bins, we measured the equivalent width of Ca II K for stars with metallicities of [Fe/H] = -2.0 through [Fe/H] = +0.5 (with bin sizes of $0.5$ dex), for the equivalent model of each spectral subtype. We calculated a 5-sigma clipped average of 
the regions to the left (4010 \AA\ --4030 \AA) and to the right (3913 \AA--3933 \AA) of the feature to fit a continuum value. Depending on the temperature of the model spectrum, we either defined a 12 \AA\ region (for stars with $T_{eff} < 8100$ K) or a 6 \AA\ region (for stars with $T_{eff} > 8100$ K) around the central wavelength of 3933.7 \AA, to completely encase the absorption line. Using the equivalent widths of the Ca II K lines measured from the model spectra, we identified a relation between metallicity and the strength of the Ca II K absorption line (Table \ref{t:Astar}). We then measured the equivalent width of the line using the same numerical integration technique on all the BOSS spectra, which were first shifted into their rest frame using the radial velocities (See Section \ref{RV}). We then separated the spectra into metallicity bins using the equivalent width regions in Table \ref{t:Astar}.

\begin{deluxetable}{c c c c c c l}
\tablecolumns{6}
\centering
\tablewidth{0pt}
\tablecaption{BT Settl A star CaII K EWs (\AA), used for metallicity bins ([Fe/H])} 

\tablehead{\colhead{Subtype} & \colhead{[Fe/H]=$-2.0$} & \colhead{$-1.5$} & \colhead{$-1.0$} & \colhead{$-0.5$} & \colhead{$0.0$}}

\startdata
                 
A3 & 0.2 & 0.35 & 0.56 & 0.74 & 0.92 \\ 
A4 & 0.36 & 0.56 & 0.87 & 1.10 & 1.50\\
A6 & 0.44 & 0.63 & 1.00 & 1.30 & 1.66\\
A7 & 0.53 & 0.84 & 1.35 & 1.71 & 2.00\\
A8 & 0.62 & 1.05 & 1.39 & 2.19 & 2.63\\
A9 & 0.74 & 1.24 & 1.97 & 2.62 & 3.09\\
\enddata

\label{t:Astar}
\end{deluxetable}

We compared our metallicity estimates to those from the SSPP as both a sanity check, and as validation that the SSPP can determine metallicity estimates passed the quoted upper temperature limit. We find that $\sim 80\%$ of the SSPP estimates are within 1 bin of our estimates. We expect this type of spread within 1 bin since even with small uncertainties in [Fe/H] ($0.1 - 0.2$ dex), the spectra can be placed in an adjacent bin. We can conclude that the SSPP and measurements of the Ca II K line are in relative agreement.  With this validation, we continue to use our derived Ca II K metallicity bins, and are confident that these cuts represent real changes in metallicity. 



\subsubsection{F, G, Early-Type K Stars}

The metallicity estimate returned by the SSPP uses a combination of 12 different methods, which include all the same spectral template fitting, neural network approaches, and photometry comparisons as mentioned in Section \ref{FGKlogg}. However, for metallicity estimates, different lines that are known to be metal sensitive (instead of gravity sensitive) are used \citep{beers99}, including the Ca II H and K lines ($\sim3950$ \AA) and the Ca II triplet ($\sim 8500$ \AA). 
After determining metallicities in all these ways, statistical outliers were removed from the sample and the final bi-weight average of the metallicity of the star is returned \citep{Lee08} in the same way as was done for the log $g$ values.  This method is accurate to $\sigma$([Fe/H]) = 0.11 dex \citep{Lee082}. 

\subsubsection{Late-Type K and Early-Type M Stars}
\label{KMfeh}

Low mass stars' molecule rich atmospheres make them extremely difficult to model \citep{allard12}, and therefore we cannot use the same method as the A stars (comparison to models) to measure their metallicities. Recent work using M dwarfs with wide binary FGK companions have led to several new techniques for estimating M dwarf metallicities. \citet{mann13} uses $\sim120$ absorption features in K5 through M4 dwarfs that are shown to be metal sensitive to estimate metallicities to a precision of $<0.1$ dex for the metallicity range $-1.04 < $[Fe/H]$ < +0.56$. We employed the \citet{mann13} method and measured metallicities for over 3,000 individual M dwarfs (no giants). Each spectrum was separated into a metallicity bin: [Fe/H] $< -1.0$; -1.0 $<$ [Fe/H] $< -0.5$; $-0.5 <$ [Fe/H] $< 0.0$; $0.0 <$ [Fe/H] $< 0.5$; and [Fe/H] $> 0.5$.

\subsubsection{Late-Type M Stars} 
\label{Mfeh}

The majority of methods for determining the metallicity for late-type M dwarfs (M5-M8) use IR spectral features
\citep[e.g., ][]{mann14, newton14}, which are not within the BOSS spectral coverage. \citet{newton14} also showed that the spread in the $J-K$ color around the stellar locus in a $J-K, H-K$ color diagram is primarily due to metallicity. We obtained $J, H,$ and $K$-band photometry by matching the SDSS catalog with the 2MASS catalog. We therefore separated the M dwarf spectra (not giants) into rough metallicity bins using IR color cuts Equation 15 in \citet{newton14}, which have a precision between $0.1$ dex and $0.5$ dex. We created three metallicity bins: [Fe/H] $<-0.5$; $-0.5 <$ [Fe/H] $< 0.0$; and [Fe/H] $> 0.0$. Even though the uncertainties are relatively high, the overlap between the high and low metallicity bins should be minimal since the difference in metallicity between our low and high metallicity bin is greater than the 0.5 dex uncertainty stated by \citet{newton14}.

\subsection{Co-adding}
\label{coadd}

The BOSS spectra are logarithmically spaced and in vacuum wavelengths. Using the radial velocities determined as described above, we shifted all of the spectra into their rest frames. We used a similar method to \citet{bochanski07}, except our wavelength grid was spaced evenly in logarithmic space (intervals of 5  km s$^{-1}$) instead of linearly spaced. The flux was then inserted (without interpolation) into the appropriate location on the flux grid corresponding to the shifted wavelength. The flux grids were then normalized to the flux at 8000 \AA. The spectral resolution is increased in this process because the shifts in radial velocity can be measured to precisions of $\sim 5-10$ km s$^{-1}$
 and were added into a grid with a spacing of 5 km s$^{-1}$. The process is essentially combining many low resolution spectra into a higher resolution template, and is very similar to the ``drizzle" process used to combine astronomical images \citep{bochanski07}.   
During the co-addition, we also corrected for a spike in flux located around 5600 \AA\  present in many of the spectra, caused by stitching the spectra from the red and blue BOSS cameras together. Finally, we trimmed the grids at 3650 \AA\ and 10200 \AA, to avoid areas that were not complete after the radial velocity shifts.
We also propagated the uncertainty reported by BOSS for each spectrum throughout this process and calculated the standard deviation of the co-added templates at each spectral channel for each template. 


\section{Results}

\subsection{The Spectra}
\label{Results:spec}

\begin{deluxetable*}{c c c c c c l}
\tablecolumns{6}
\centering
\tablewidth{0pt}
\tablecaption{The Template Spectra} 

\tablehead{\colhead{File Name} & \colhead{Number of Co-Added Spectra} & \colhead{CaK} & \colhead{NaD\tablenotemark{1}} & \colhead{Halpha}& \colhead{TiO8440} }

\startdata
                 
O5 & 10 & 1.015 & 1.013 & 0.892 &  0.912\\ 
B6 & 5 & 0.986 & 1.019 & 0.834 & 1.029\\
A6\_-0.5\_Dwarf & 3 & 0.953 & 1.003 & 0.788 & 0.979 \\
F6\_+0.0\_Dwarf & 29 & 0.648 & 0.965 & 0.881 & 0.985 \\
G6\_-1.5\_Giant & 13 & 0.587 & 0.975 & 0.922 & 0.987 \\
K5\_-0.5\_Dwarf & 11 & 0.588 & 0.781 & 0.954 & 0.991 \\
M4\_+0.5\_Dwarf & 7 & 1.010  & 0.672  & 1.053  & 0.902 \\
L1 & 34 & 0.613  & 1.391  & 1.043  &  0.683 \\

\enddata
\tablenotetext{1}{\footnotesize spectral indeces for any doublets (e.g., Na I D) are for the blended feature and therefore include both lines}

\label{t:Spectra}
\end{deluxetable*}

\begin{figure*}[]
\begin{center}
\includegraphics[scale=0.7]{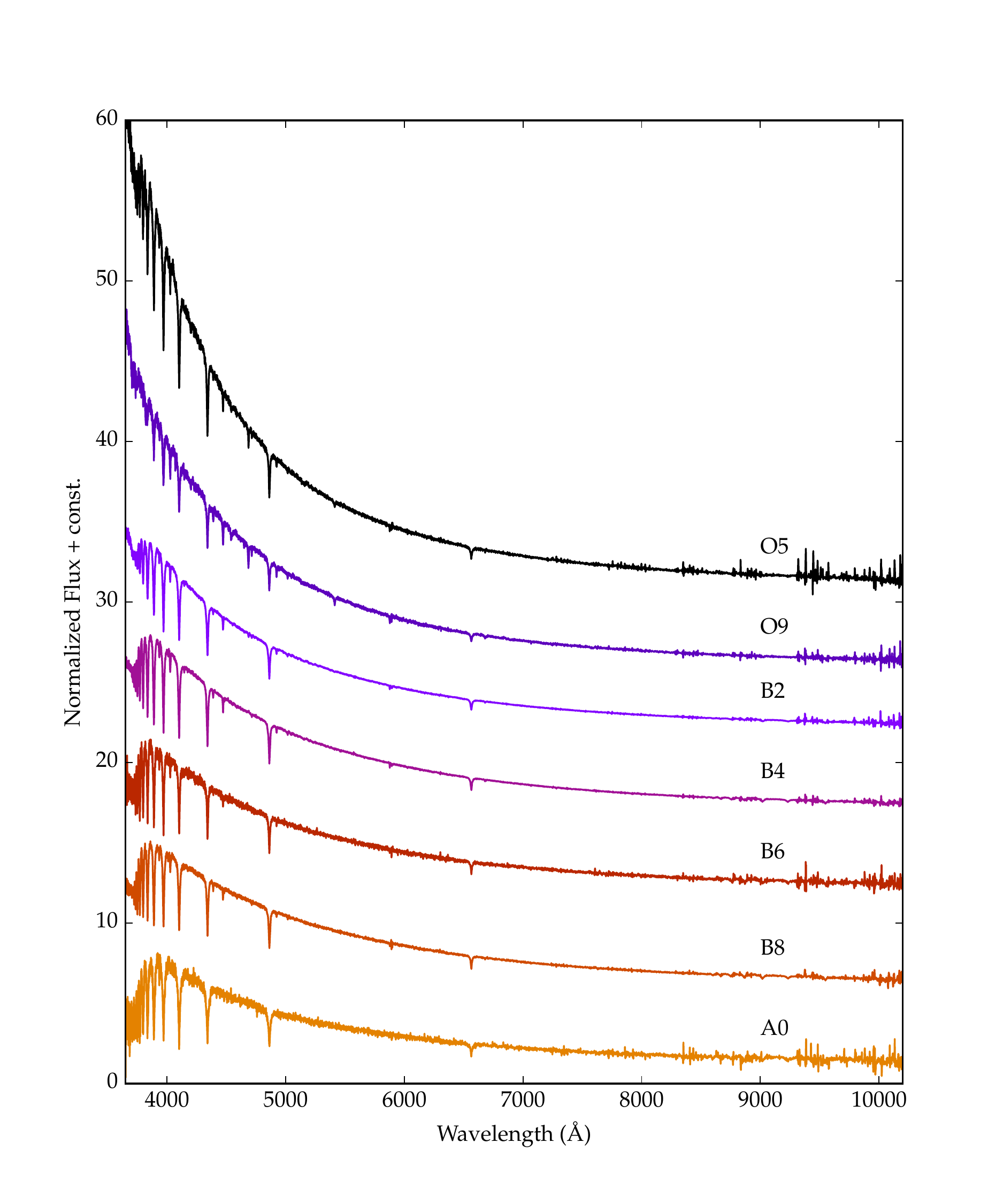}
\caption{\small
Sample of the template spectra for high-mass spectral types. The spectra are all normalized at 8000 \AA, and a constant is added to each template to improve readability. 
All of the high-temperature templates are available in FITS format in the online journal}
\label{f:OBAspec}
\end{center}
\end{figure*}

\begin{figure*}[]
\begin{center}
\includegraphics[scale=0.7]{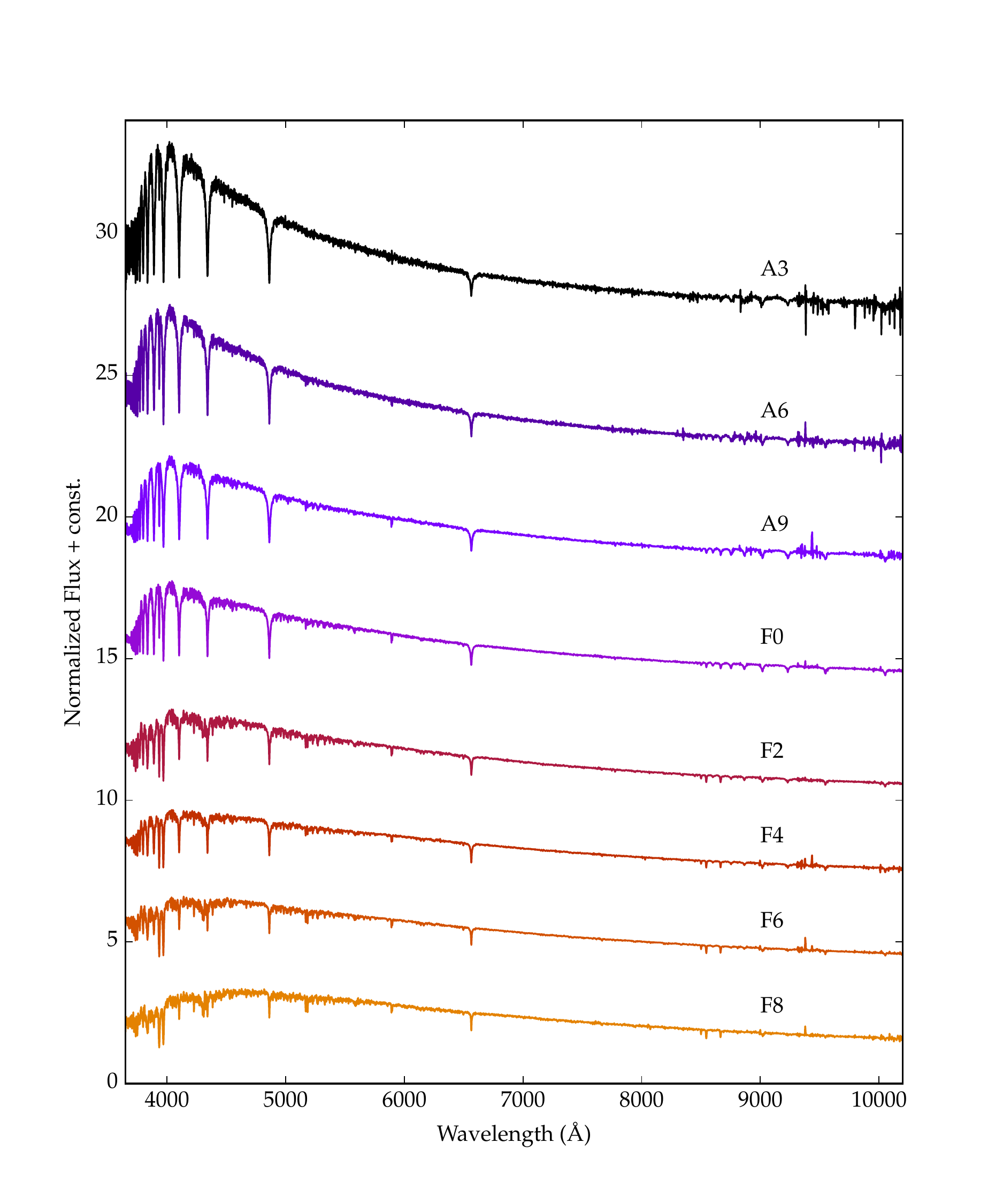}
\caption{\small
Sample of the template spectra for A and F spectral types, at solar metallicity ([Fe/H] = 0.0). The spectra are all normalized at 8000 \AA, and a constant is added to each template to improve readability.
All of the A and F-type templates are available in FITS format in the online journal}
\label{f:AFspec}
\end{center}
\end{figure*}

\begin{figure*}[]
\begin{center}
\includegraphics[scale=0.7]{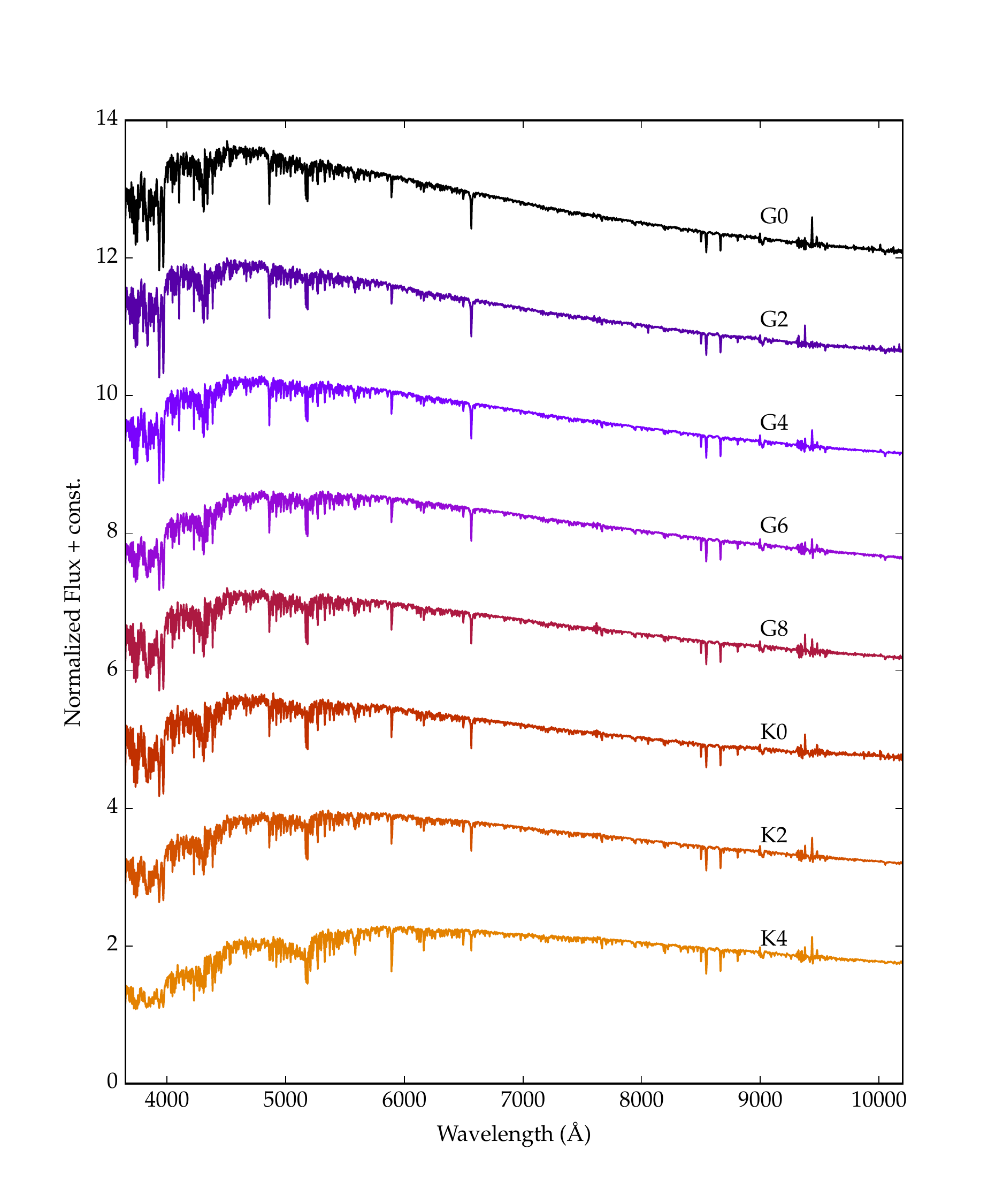}
\caption{\small
Sample of the template spectra for the main-sequence, solar-mass spectral types at solar metallicity ([Fe/H] = 0.0). Again, the spectra are all normalized at 8000 \AA, and a constant is added to each template to improve readability. 
All of the G and K-type templates are available in FITS format in the online journal}
\label{f:GKspec}
\end{center}
\end{figure*}

\begin{figure*}[]
\begin{center}
\includegraphics[scale=0.7]{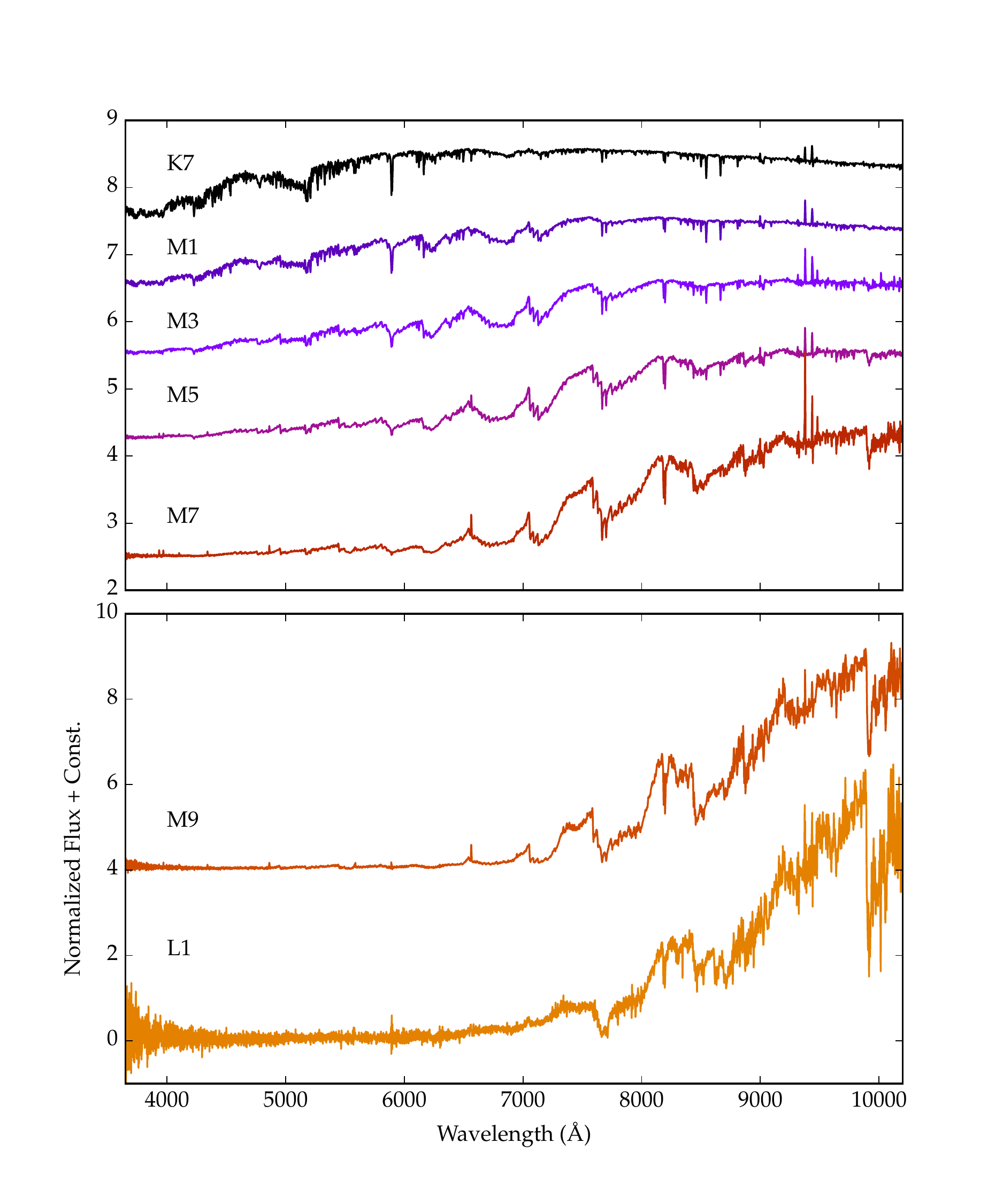}
\caption{\small
Sample of the template spectra for the main-sequence, low-mass spectral types at solar metallicity ([Fe/H] = 0.0). The spectra are all normalized at 8000 \AA, and a constant is added to each template to improve readability. The late-type M and L stars have different scales because most of their flux is concentrated red-ward of 8000 \AA, where the normalization occurs. We therefore put them on a separate set of axes, so they would not dominate the other low-mass stars. 
All of the low-temperature templates are available in FITS format in the online journal}
\label{f:MKspec}
\end{center}
\end{figure*}

\begin{figure*}[]
\begin{center}
\includegraphics[scale=0.55]{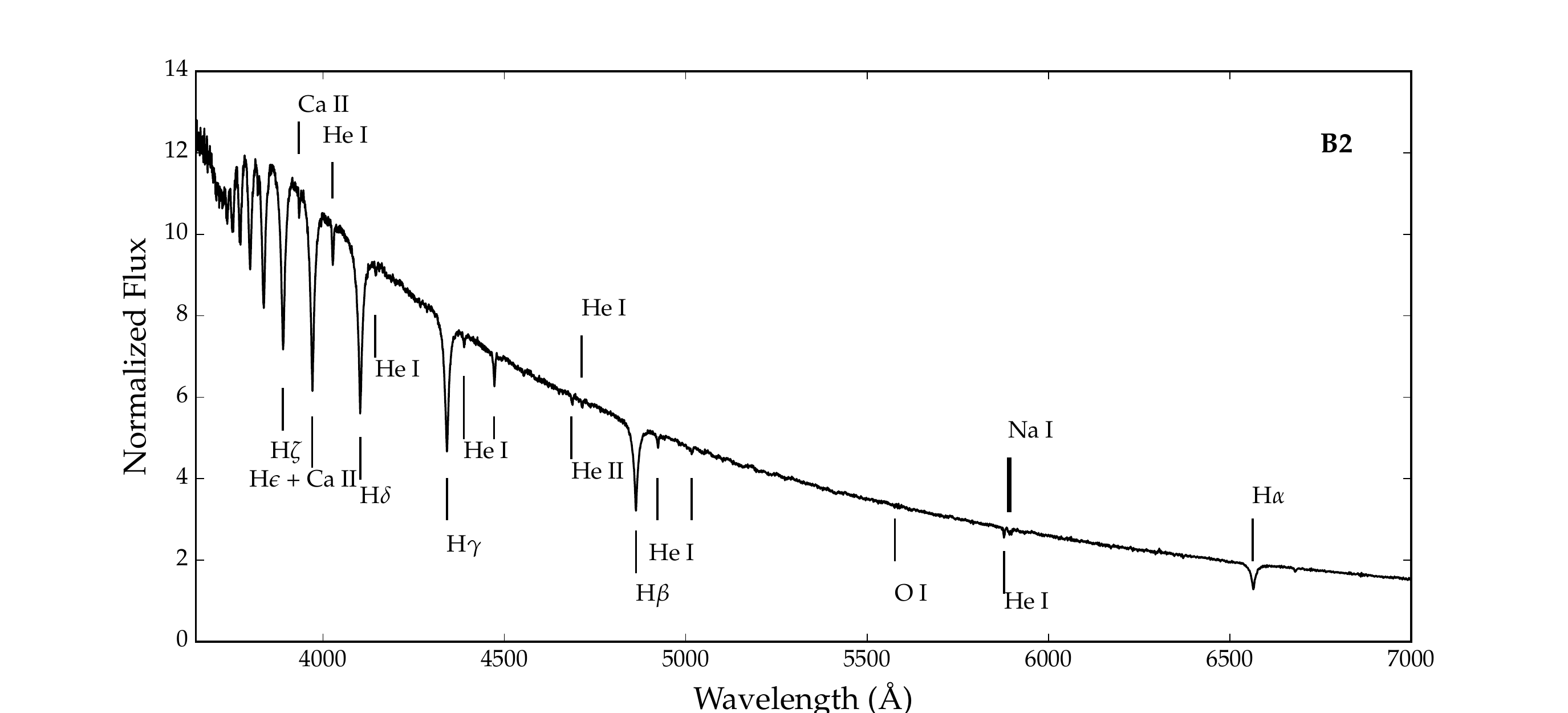}

\includegraphics[scale=0.55]{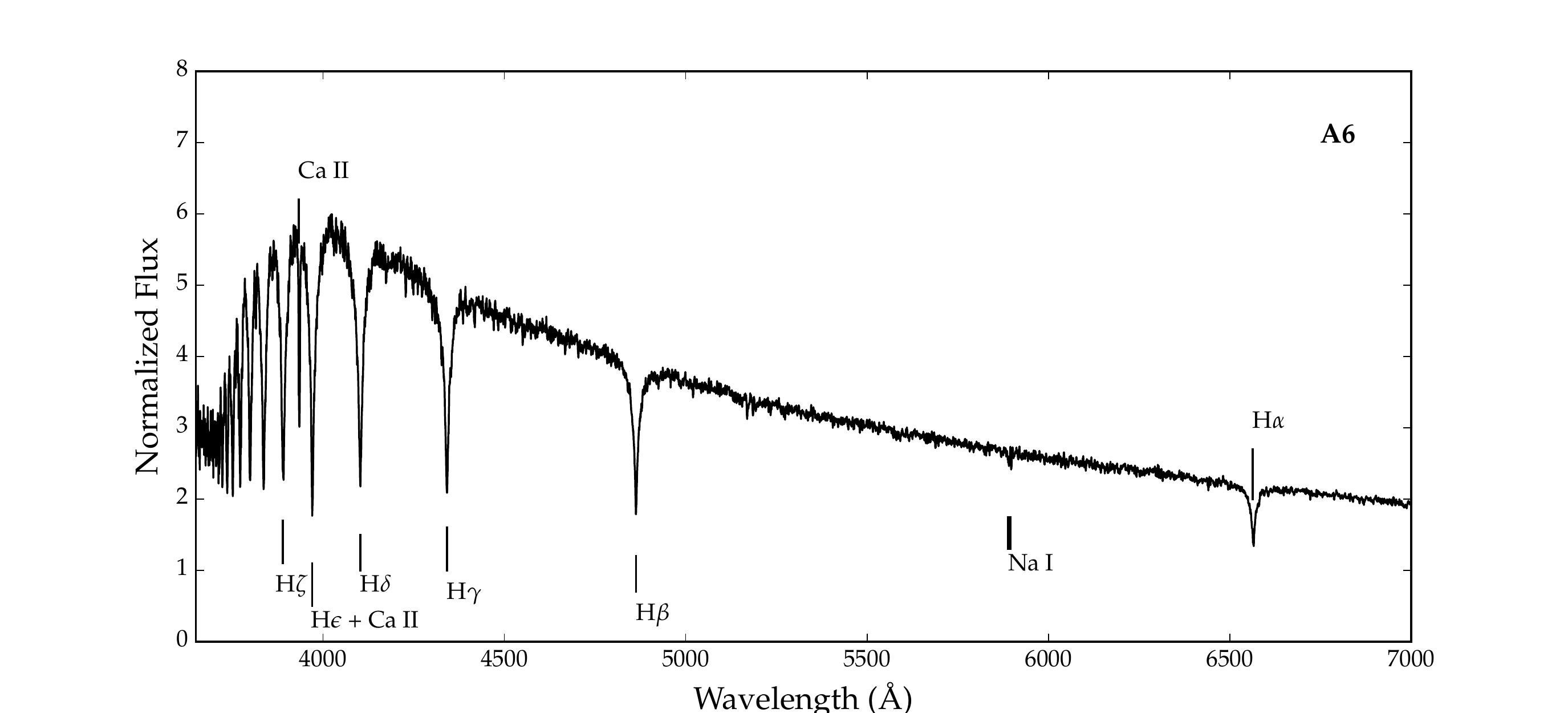}

\includegraphics[scale=0.55]{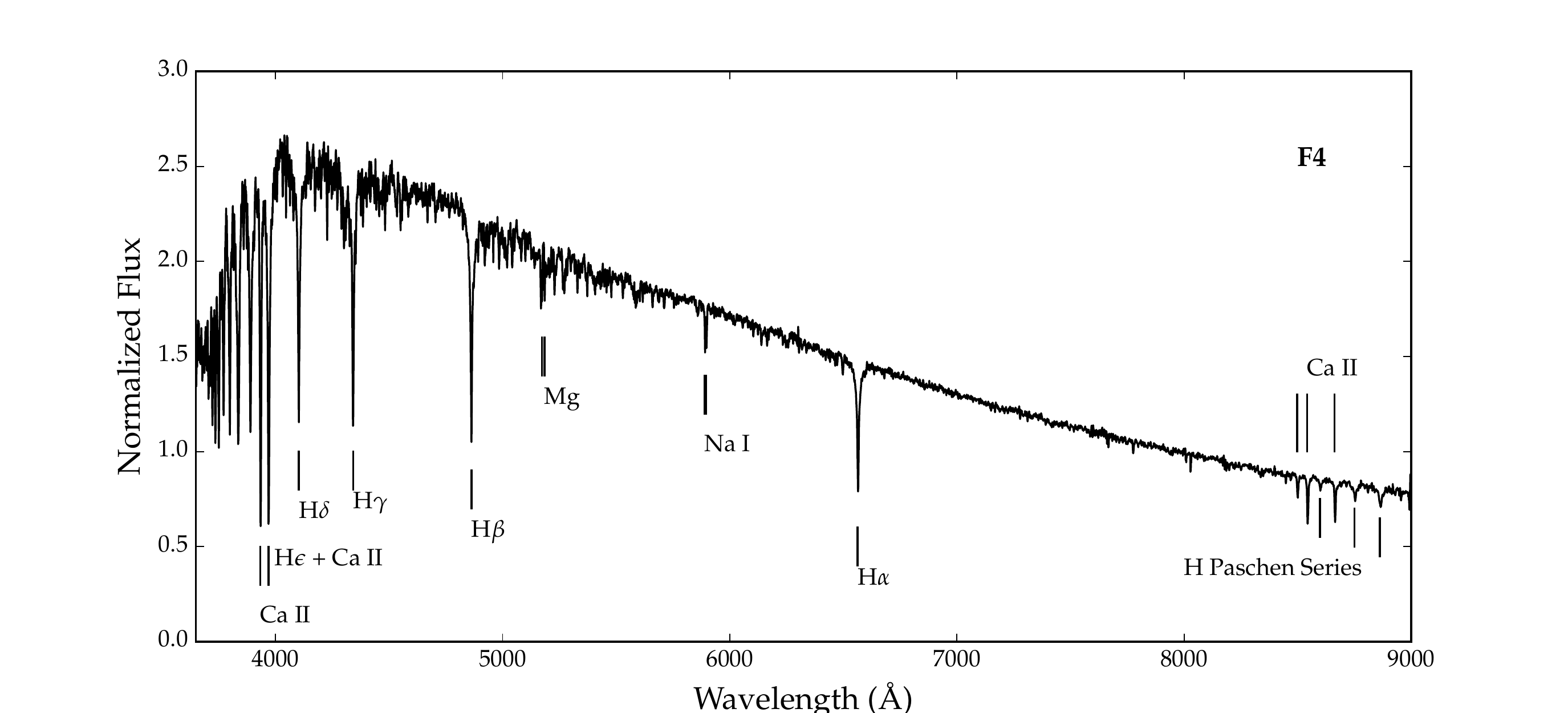}

\caption{\small
Example spectra from main-sequence, high-mass spectral subclasses (B2, A6, F4) with solar metallicity when applicable ([Fe/H] = 0.0). All the prominent absorption features are labeled for each spectrum.}
\label{f:label_highmass}
\end{center}
\end{figure*}

\begin{figure*}[]
\begin{center}
\includegraphics[scale=0.55]{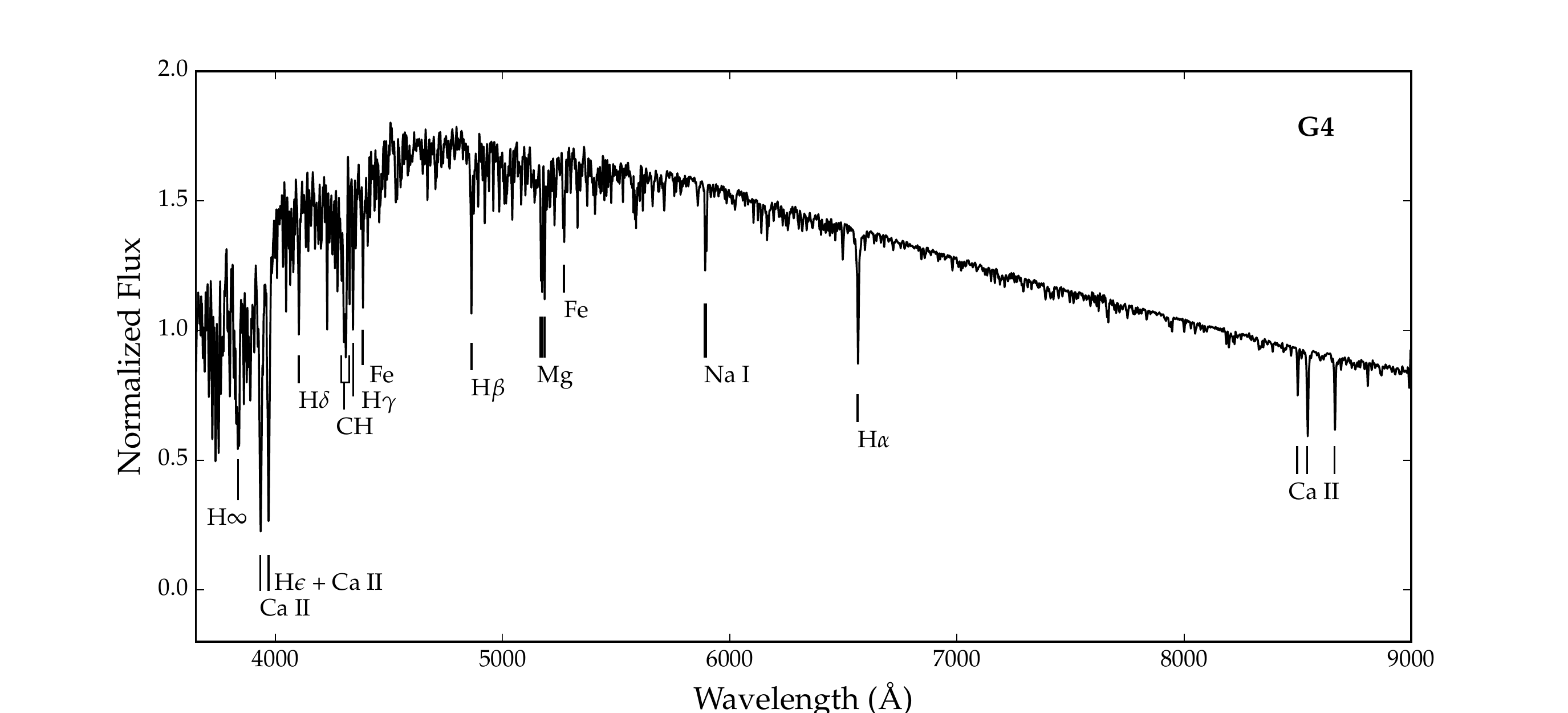}

\includegraphics[scale=0.55]{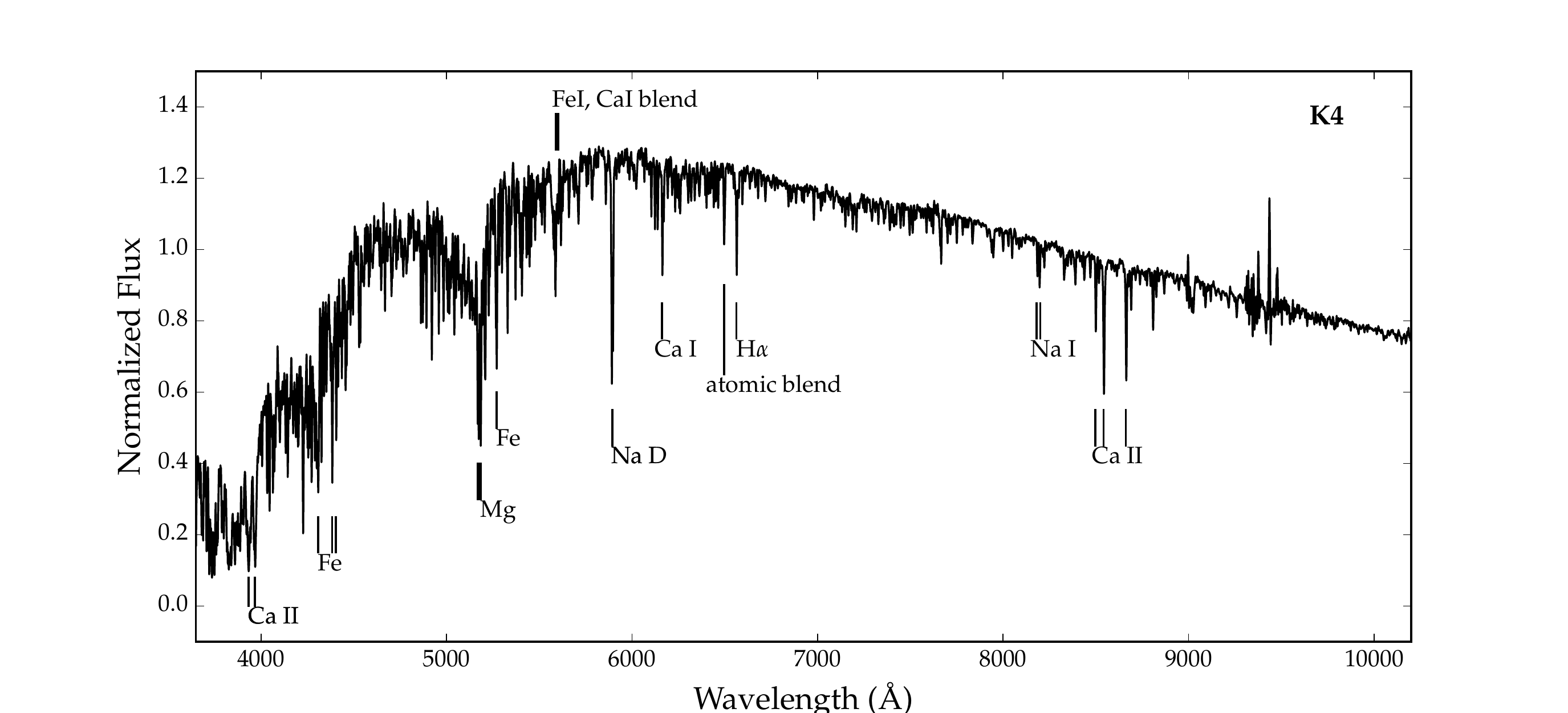}

\includegraphics[scale=0.55]{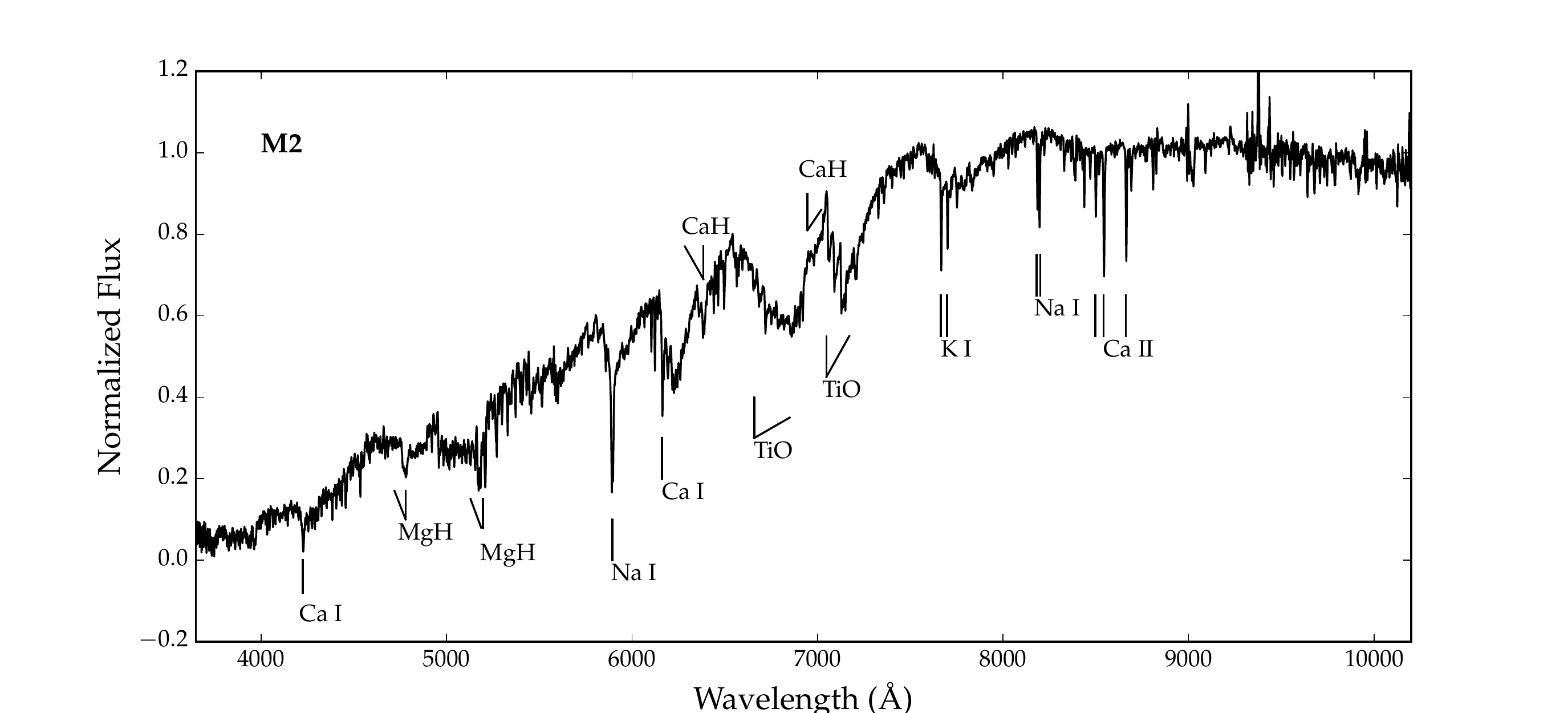}

\caption{\small
Example spectra from main-sequence, low-mass spectral subclasses (G4, K4, M2) with solar metallicity ([Fe/H] = 0.0).  All the prominent absorption features are labeled for each spectrum. The `V' shapes are used instead of lines to show large molecular bands (in the spectrum of the M2). 
}
\label{f:label_lowmass}
\end{center}
\end{figure*}

We present a subset of our 324 template spectra in Table \ref{t:Spectra}; the full table is available for download in the online journal. Each entry displays the number of BOSS spectra co-added to create the template, along with measurements of 29 spectral indices. The spectral indices, which are ratios of the flux in the region of the absorption feature divided by the continuum flux, are the features used by PyHammer to estimate the spectral type and metallicity (see Section \ref{pyhammerEstimates} for details on the spectral indices). 
Figures \ref{f:OBAspec} through \ref{f:MKspec} display a subset of the template spectra. All of the spectra are available in fits formats by clicking on the link for the data behind the figures, and online in SDSS formatted fits binary tables \footnote[2]{https://doi.org/10.5281/zenodo.321394}, or in standard fits format in a public DropBox folder \footnote[3]{https://www.dropbox.com/sh/3qo9fsxn6vegggg/AACa6-9CmAW5ovPvRpQ4vE9Da?dl=0}. For those interested in the complete sample, all the individual BOSS spectra co-added for each template are available in a Zenodo repository \footnote[4]{https://doi.org/10.5281/zenodo.344471}, organized by the spectral type, luminosity class, and metallicity, or by contacting the corresponding author. Figures \ref{f:label_highmass} and \ref{f:label_lowmass} show one spectrum from each spectral class, with all of the prominent absorption features labeled.  These figures will be especially useful in the classroom.  The templates and associated figures can be used by introductory students first viewing stellar spectra, as well as the seasoned astronomer, who may not have seen high-quality spectra that span the entire optical wavelength and parameter space.

\begin{figure*}[]
\begin{center}
\includegraphics[scale=0.5]{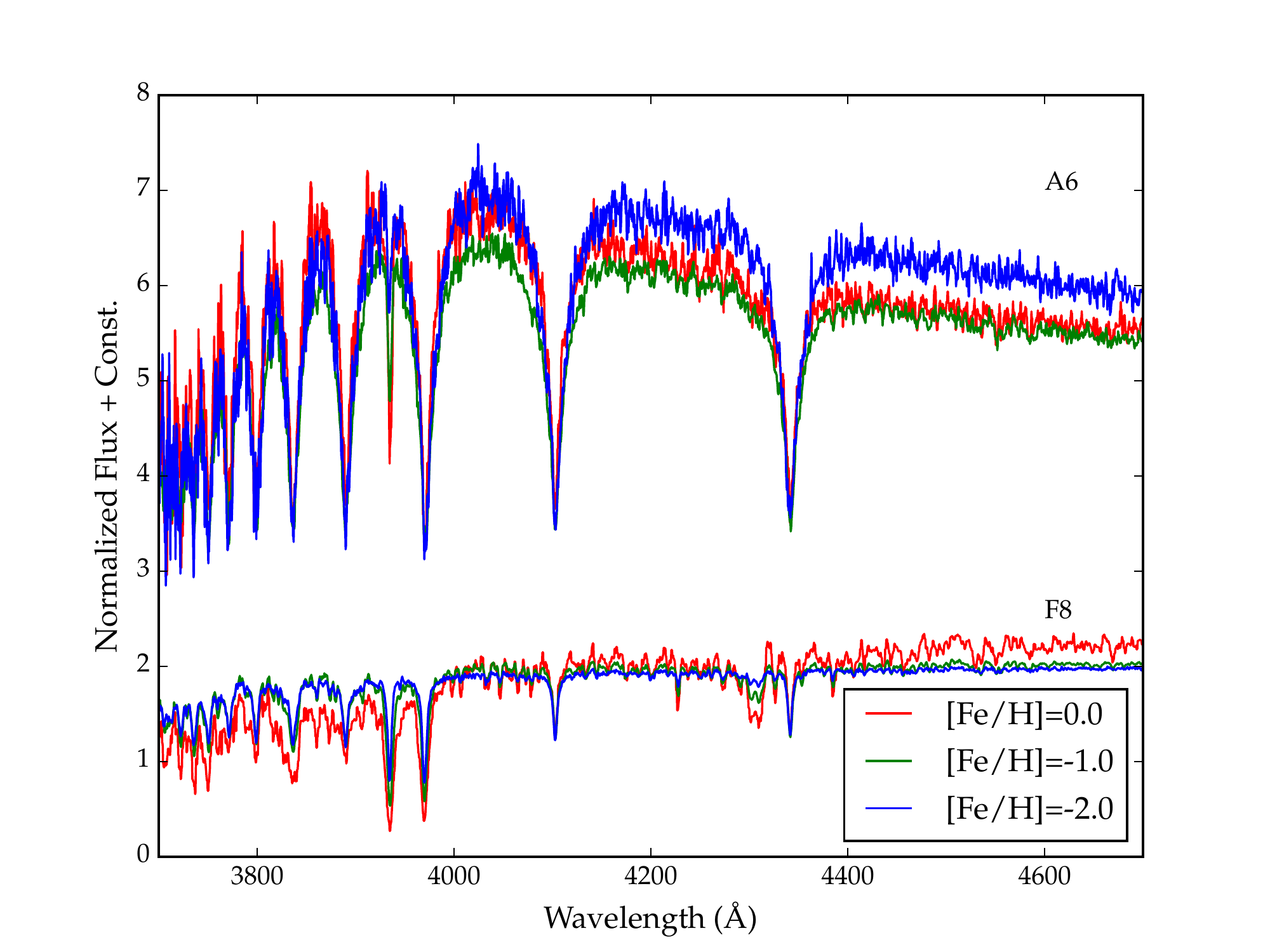}

\includegraphics[scale=0.5]{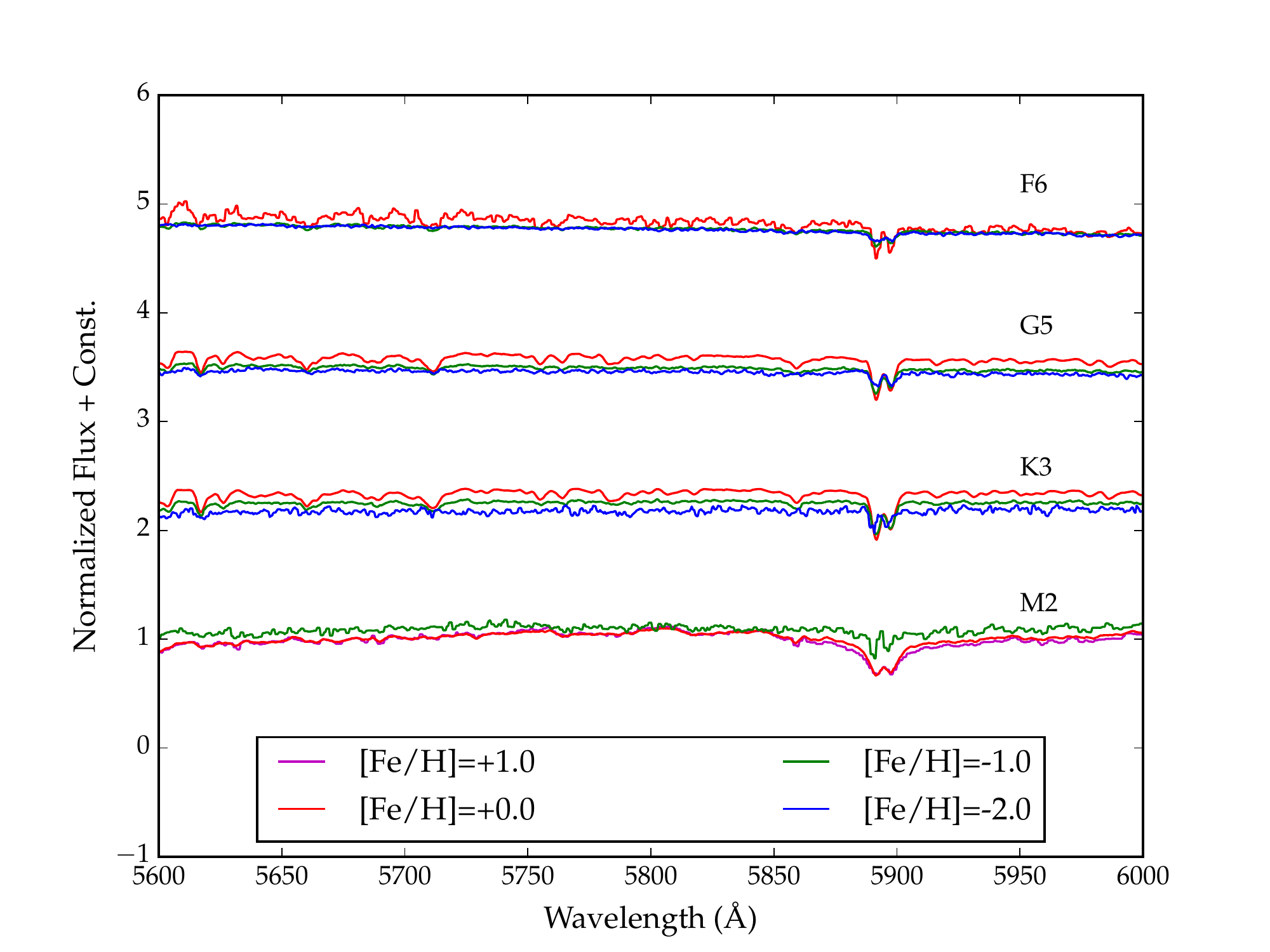}
\caption{\small
Metallicity variations for a range of spectral subtypes, shown in two different wavelength regions. The top panel shows the region around the Ca II H and K lines ($\sim3940$ \AA), which are metallicity sensitive for the higher temperature stars (A6, F8). The bottom panel shows the region around the sodium doublet ($\sim5900$ \AA), which is metallicity sensitive for the spectral subclasses around F through early M-type stars. The equivalent widths of this particular sodium doublet (Na I D) are smaller for late-type M stars, making it difficult to use the feature to distinguish metallicity. The templates of the same spectral classes have no added constants to the flux, but the templates of different spectral classes have an added constant for readability. 
}
\label{f:metalSpec}
\end{center}
\end{figure*}

Figure \ref{f:metalSpec} shows an expanded view of the Ca II K and Na I D metal sensitive features in different regions of the spectra (3933 and 5890 \AA, respectively). The Ca II K line is prominent in high-temperature stars, while the Na I D lines are almost absent. In low-temperature stars, the trend is the opposite. We find that the Na I D lines are a useful metallicity indicator for F through early M-type stars, and the Ca II K line to be a useful metallicity indicator for A and F stars. The higher metallicity spectra have deeper absorption features (larger equivalent widths) for both of these features. In the bottom panel of Figure \ref{f:metalSpec}, the lowest metallicity templates also show very few features outside of the neutral sodium D doublet and are virtually flat for the F6 and G5 spectra. The K3 and M1 spectra have very few low-metallicity spectra that are co-added to create the templates, and are therefore more noisy than their higher metallicity counterparts, yet still show a lack of real absorption features. 

\begin{figure*}[]
\begin{center}
\includegraphics[scale=0.65]{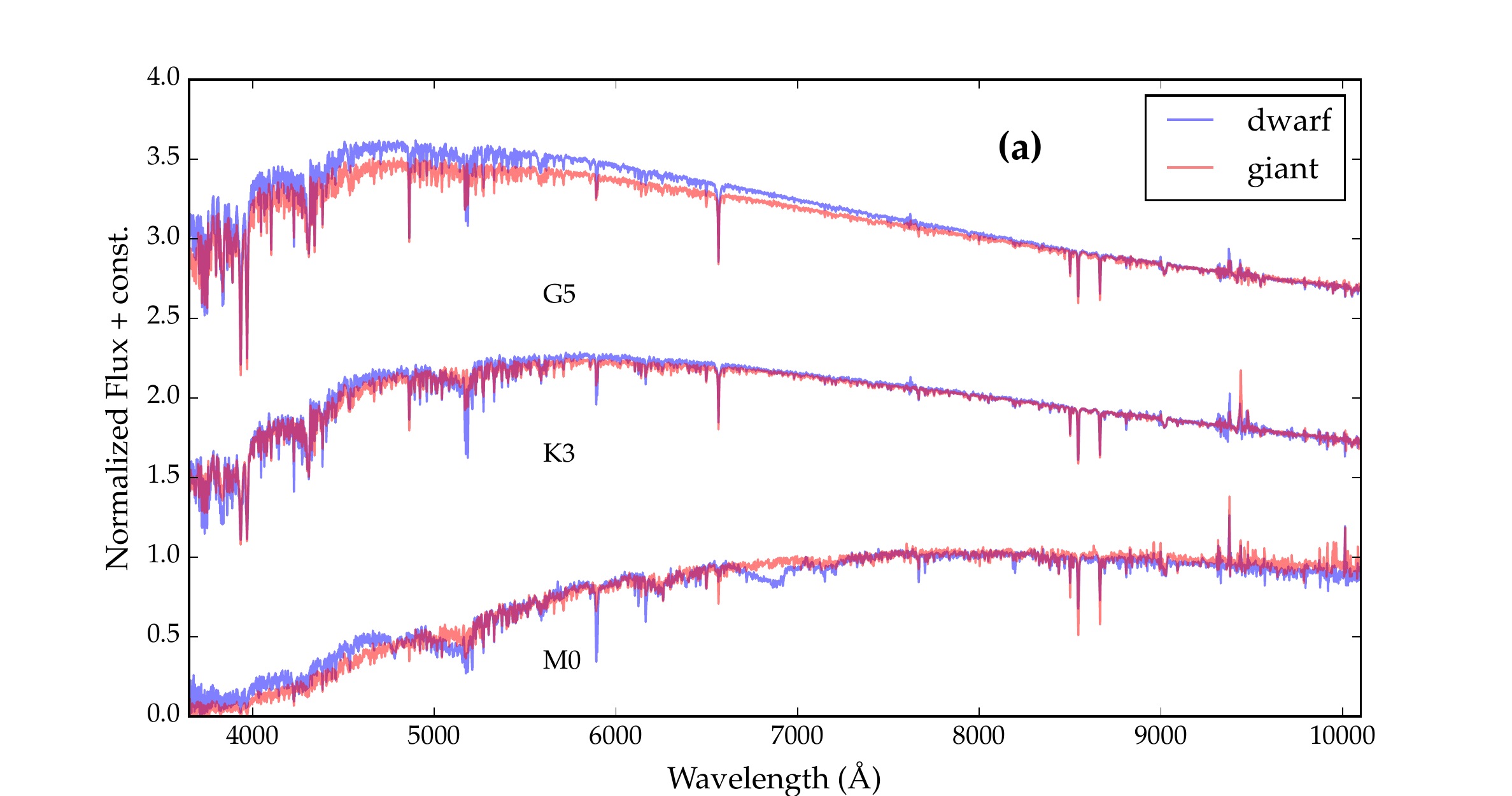}
\includegraphics[scale=0.65]{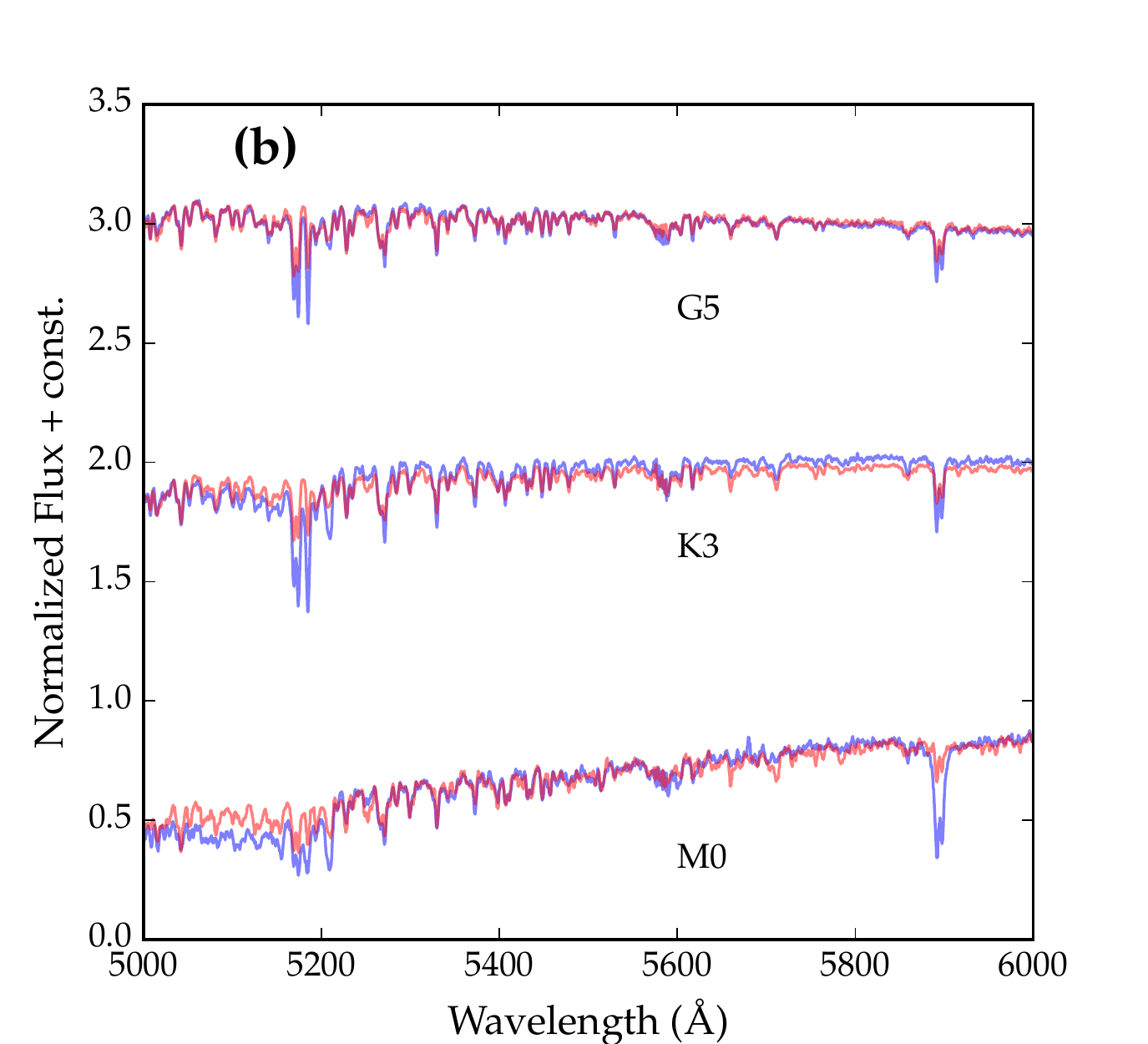}
\includegraphics[scale=0.65]{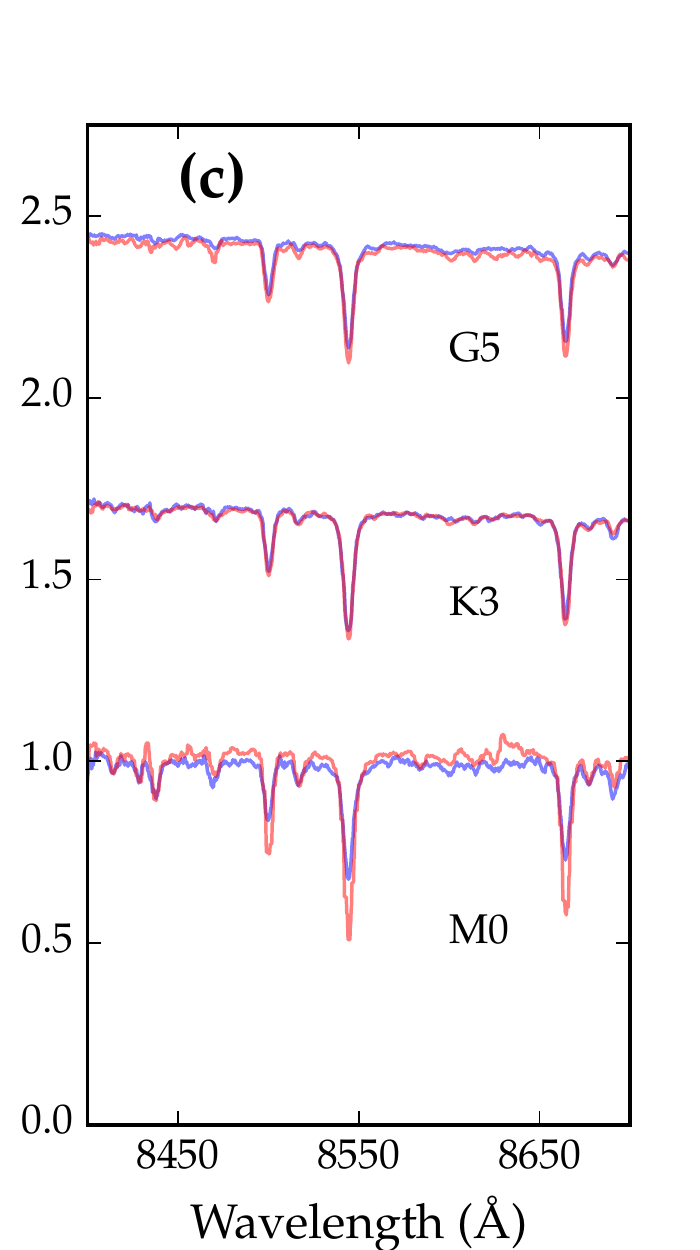}
\caption{\small
Surface gravity comparison between dwarf and giant templates of the same metallicity ([Fe/H] = -1.0 for the G5 and K3, [Fe/H] = -0.5 for the M0 dwarf, unknown metallicity for M0 giant). The red line shows the giant template, and the blue line shows the dwarf template. 
Panel \textbf{(a)} shows the entire spectrum, while panel \textbf{(b)} shows the expanded region around the Na I D lines (5900\AA) and Mg b/MgH feature (5200\AA), and panel \textbf{(c)} shows the zoomed in region around the Ca II triplet. All of these lines are known to be sensitive to surface gravity, which is confirmed in our templates. 
}
\label{f:dwarfgiantSpec}
\end{center}
\end{figure*}

Figure \ref{f:dwarfgiantSpec} is similar to Figure \ref{f:metalSpec} except it shows differences between dwarf and giant spectra for a range of spectral types. Because G and K stars were selected with the SSPP, we have metallicities for both the dwarf and the giant star, and therefore show the same metallicity and spectral type. 
The top panel of Figure \ref{f:dwarfgiantSpec} shows the entire spectrum for both a dwarf and a giant spectrum for the spectral classes G5, K3 and M0. The continua (even without features) are slightly different, with more flux in giants coming from the red end, and more flux in the dwarfs coming from the blue end of the spectrum. The bottom panels show a closer look at three absorption features, which which are sensitive to surface gravity (bottom left: Mg b/MgH and Na I D, bottom right: Ca II triplet). The Mg b and MgH feature around 5100 \AA\ is absent in metal poor giants, weaker in Population I giants as compared to dwarfs, and prominent in dwarfs \citep{helmi03}.
The M giant does not include a metallicity estimate, but because the Mg/MgH feature is absent, we can assume it is a sub-solar metallicity giant. 
The Na I D lines are also often used as an age indicator since the feature is extremely strong in dwarfs and nearly non-existent in giants \citep{schlieder12}. Alkali atoms in general are all extremely sensitive to density because the sole valence electron can easily be perturbed by small changes in pressure \citep{schlieder12}, making sodium transitions good surface gravity (and age) indicators. Lastly, the equivalent width of the calcium triplet around 8600 \AA\ is known to be larger in giants than in dwarfs \citep{jones84}, which our spectra show in the the bottom left panel. 
 
In our sample, we find low-metallicity K giants, but no high-metallicity K giants or late-type M giants (of any metallicity). Stellar evolution models show that low and high-metallicity giants occupy two different temperature regimes. Stars with lower metallicity become hotter giants since they cannot cool as efficiently, while higher metallicity giants can cool more quickly and therefore occupy the late-type M giant regime \citep{girardi04, bressan12}.
With the stellar evolution models in mind, 
the lack of late-type giants is expected in the SDSS data; there is a strong bias in the metallicity of the giants we can observe with SDSS. 
Because of the sight lines observed in the SDSS footprint, the most distant stars (i.e. giants) are located at high Galactic latitudes. The stars located at large distances above the Galactic plane are on average between 0.5 and 1.0 dex more metal poor than their closer counterparts \citep{west11}. 
Therefore, in SDSS we do not expect to see many (high metallicity) M-type giants, which is validated by \citet{covey08} who find that less than 2\% of stars redder than a spectral type of K5 in a magnitude limited SDSS field were giants. 

\begin{figure*}[]
\begin{center}
\includegraphics[scale=0.7]{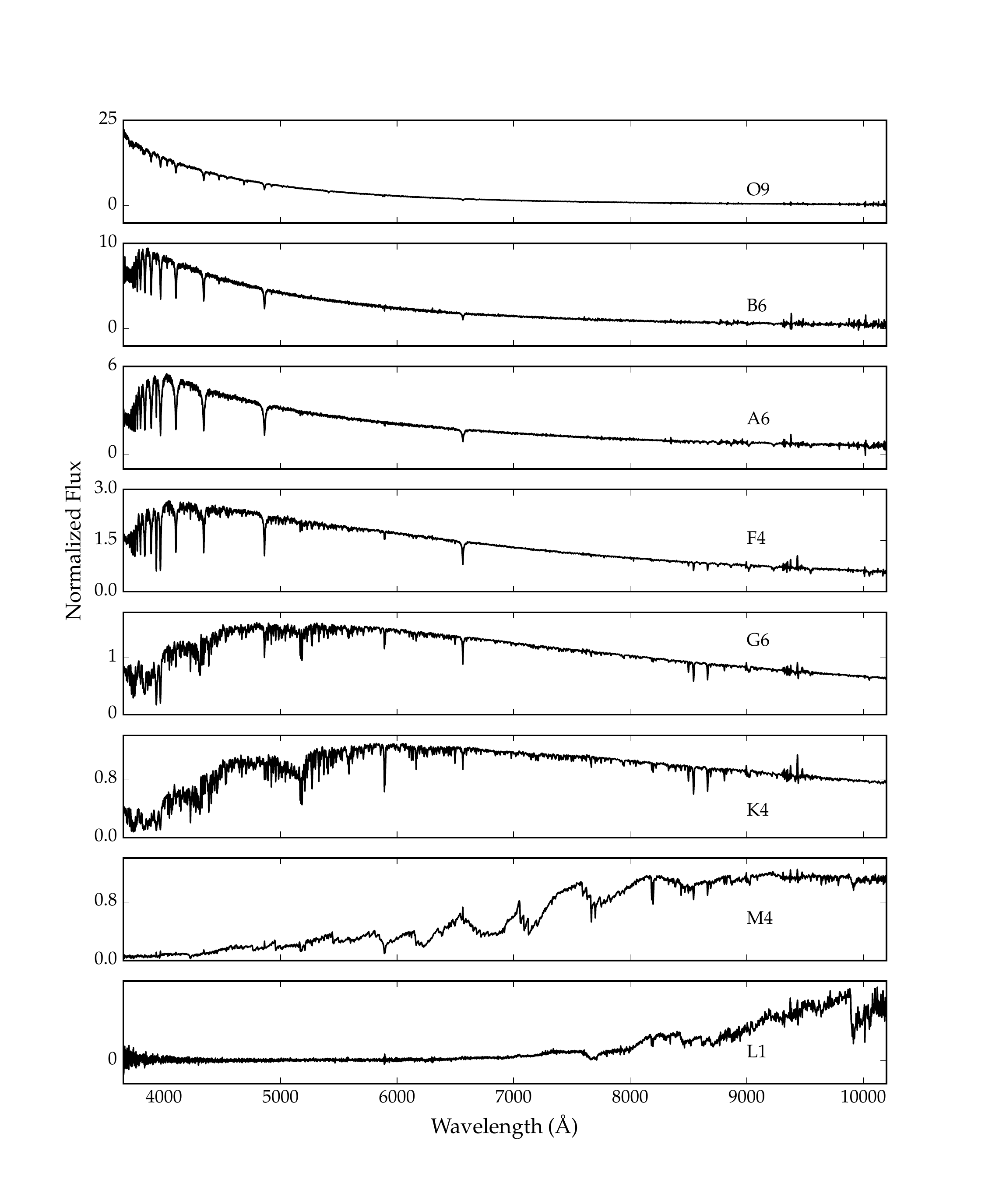}
\caption{\small
Example of one spectrum per spectral type, where each spectrum is at solar metallicity (if applicable) and classified as a dwarf (if applicable). The spectra are all normalized at 8000 \AA. 
}
\label{f:allspec}
\end{center}
\end{figure*}

\subsubsection{PyHammer}

Along with these templates, we are releasing a computing product (dubbed ``PyHammer"), which can assign both an automatic estimate of spectral type and metallicity, and/or be used to visually classify spectra. PyHammer uses the templates from this paper as comparisons to automatically determine the spectral type and an estimate of the metallicity by measuring prominent line indices and performing a weighted least squares minimization. The GUI then allows the user to visually compare their spectra to the templates and visually determine the spectral type and metallicity. This code is based on the ``Hammer" spectral typing facility \citep{covey07}, but has been updated to include metallicity information and now is in python (instead of IDL). We have also added a component to determine the radial velocity of the stars, and improved on the automatic guessing method. The PyHammer code is unique because it does not use synthetic spectra to determine metallicities and spectral types, and both has an automatic and visual or ``by-eye" component. Details of the code (including how to download and operate it) can be found in the Appendix.  

\subsection{Photometry}
\label{Results:phot}

Along with the spectra, we report average photometric colors for each template. 
Table \ref{table:photometry} shows a sample of our templates with the photometry information. The full table, which includes each template with all of the color information and associated uncertainties is available to download in ascii format in the online journal. The colors are averaged over each individual spectrum co-added in a template, after those with bad photometry are removed (i.e., flagged for being deblended, containing a cosmic ray, or saturated). The RMS is the standard deviation of all of the colors for the individual spectra, and the $\sigma$ is the propagated uncertainty in photometry provided by SDSS. 
On average, the low metallicity templates have bluer colors than the high metallicity templates. A comparison with \citet{covey07} indicates that the colors are comparable. 

\begin{figure*}[]
\begin{center}
\includegraphics[scale=0.65]{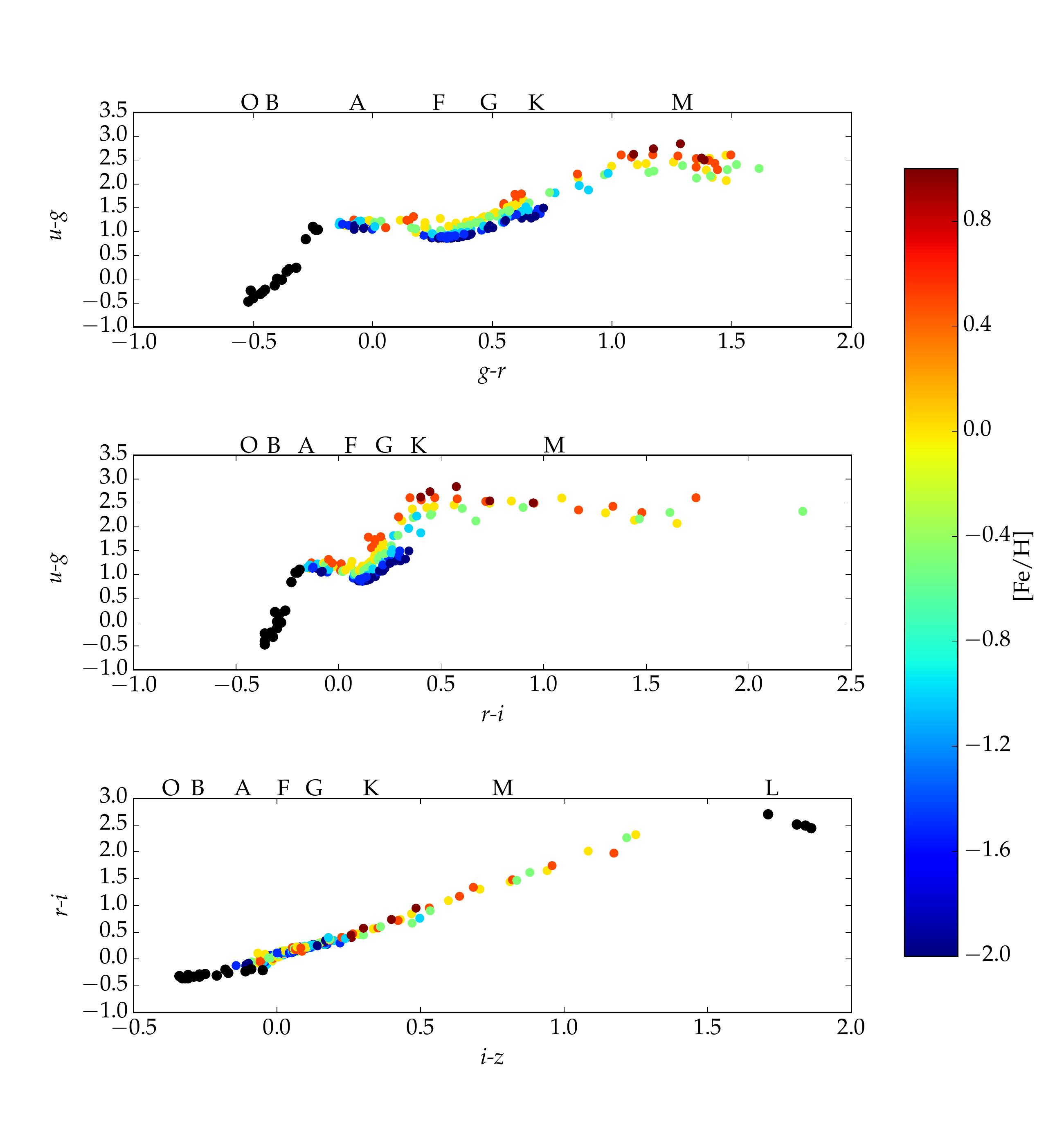}
\caption{\small
Color-color diagrams for the photometry of all of the main sequence templates. Each template is colored by its metallicity, except for the O5-A2 and L star templates, which are not separated by metallicity bins, and are colored black. A comparison with \citet{covey07} shows very similar results. There is a clear trend with metallicity, especially in the top two plots, where lower metallicity templates show bluer colors on average compared to the high-metallicity templates of the same spectral type. We only show the three color space combinations that display the most separation among metallicity bins. 
}
\label{f:colorcolor_metal}
\end{center}
\end{figure*}

Figure \ref{f:colorcolor_metal} shows graphically the trend that low-metallicity stars are bluer on average. The overall trends (ignoring metallicity) are almost identical to those shown in \citet{covey07}, however when metallicity information is added, the spread in the main sequence seems to be almost entirely due to metallicity. The top two plots in Figure \ref{f:colorcolor_metal} show an especially significant distinction between the low and high-metallicity stars, which demonstrates that the \textit{u-g} color is extremely useful for differentiating metallicities. This trend seems to be valid from F dwarfs through mid-type M dwarfs ($\sim$M4). The A stars seem to hold to the same trend, however the photometry is not as spread out in that region of color-space, making it extremely difficult to accurately distinguish metallicities. Beyond the mid-type M dwarfs, the light emitted in the \textit{u}-band is too faint to report an accurate \textit{u}-band measurement. Other studies of stars in SDSS \citep[e.g., ][]{ivezic08a} have observed color trends with metallicity in the past, however metallicity information for low-mass stars was not available for those studies. 

\begin{figure*}[]
\begin{center}
\includegraphics[scale=0.55]{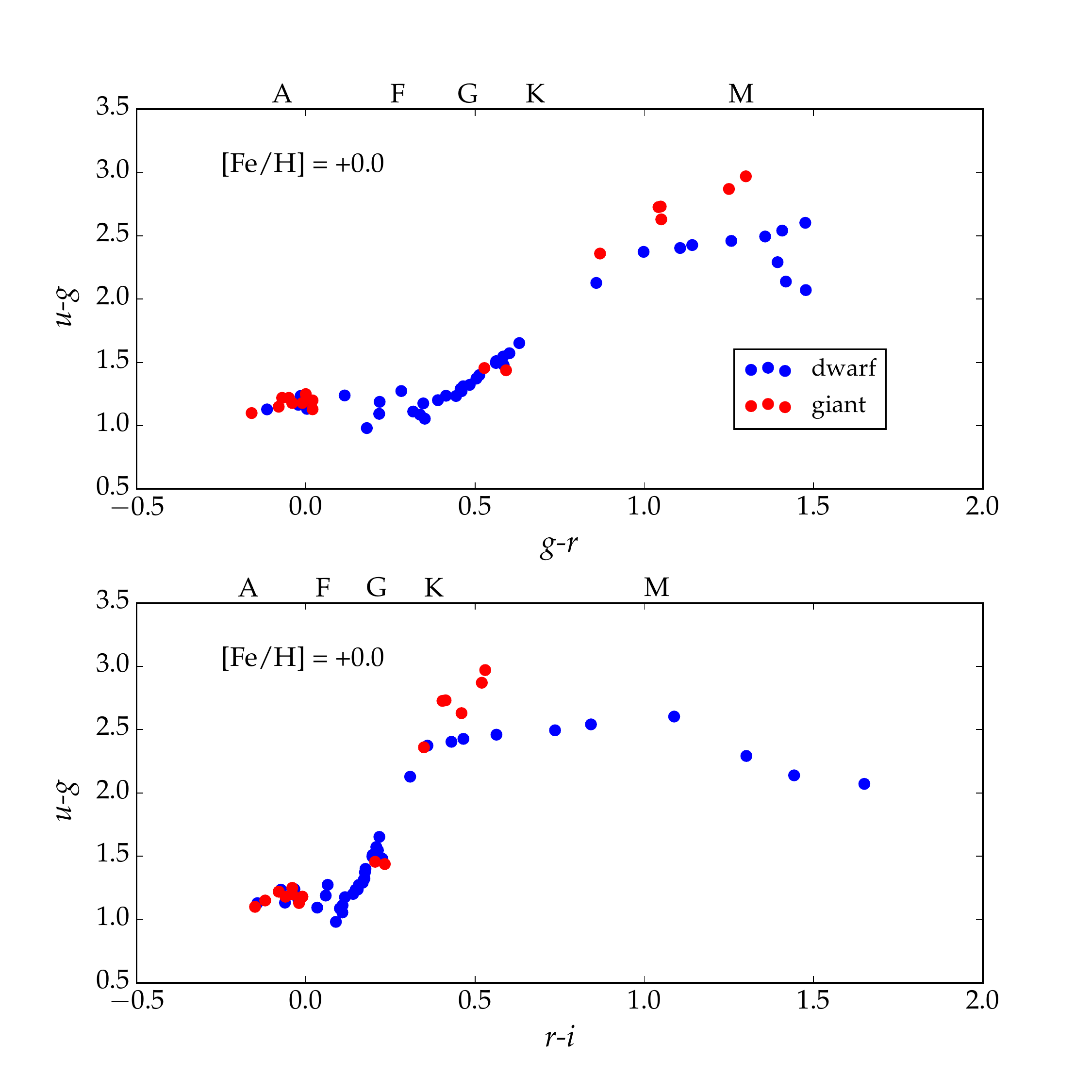}
\includegraphics[scale=0.55]{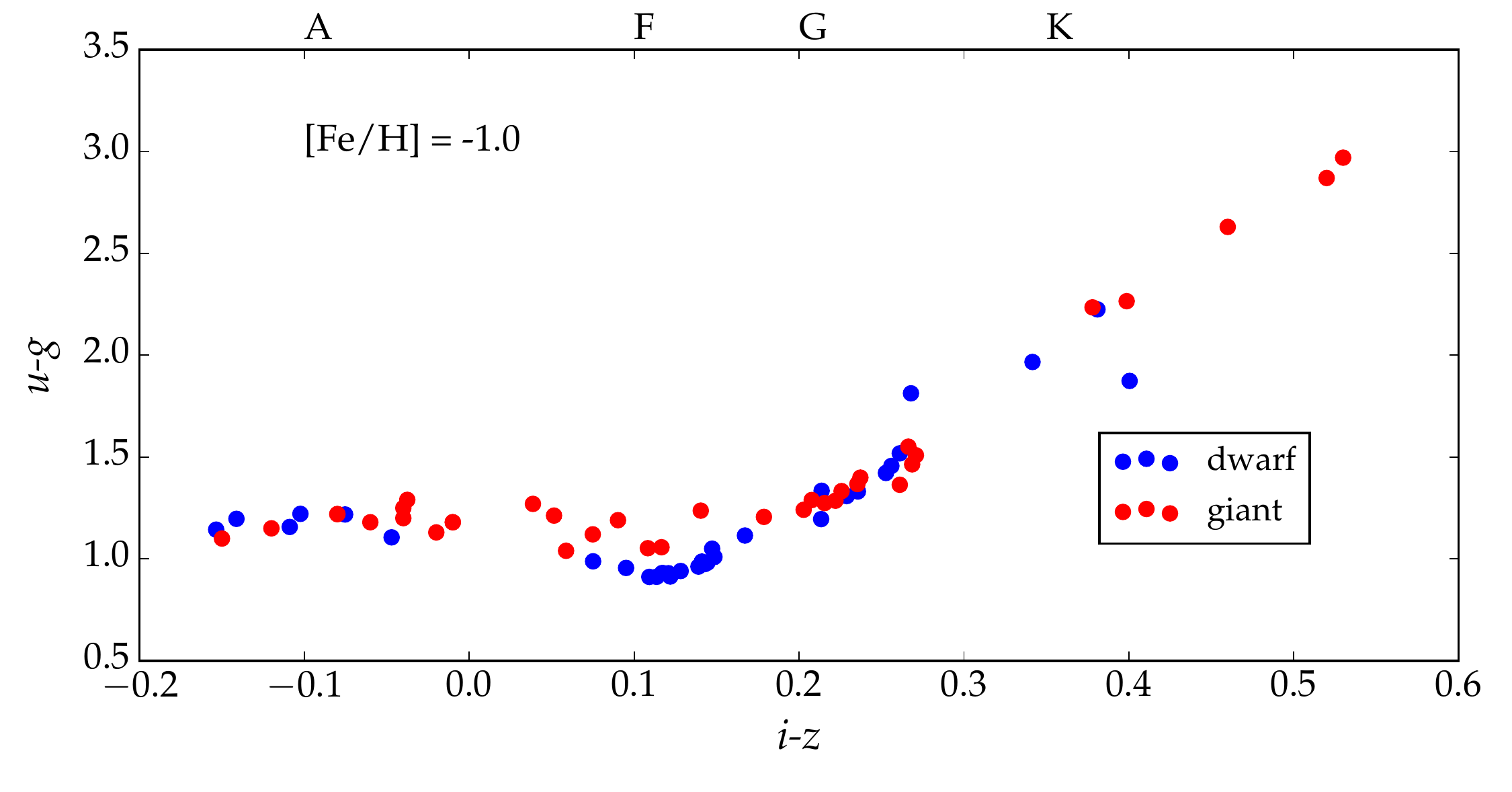}
\caption{\small
Color-color diagrams showing the main sequence (dwarf; blue) versus giant (red) luminosity classes. To compare the two accurately we only show a single metallicity bin in each plot (top two: [Fe/H] = 0.0, bottom: [Fe/H] = -1.0). Some of the giants (A-type and M-type) are not separated by metallicity and are shown in all figures. The giants show significant deviation from the dwarfs as we move to spectral types later than mid-K for all metallicity bins available. In the bottom plot, the F stars show a similar trend in the \textit{u-g} vs \textit{i-z} color-space, and are noticeably separated from the dwarfs. We do not see any clear separation in color-space for the G or A giants, however, and they are indistinguishable from the dwarfs in all our plots. 
}
\label{f:colorcolor_dg}
\end{center}
\end{figure*}

Figure \ref{f:colorcolor_dg} shows three color-color diagrams demonstrating the color difference between luminosity classes. In general, all of the plots show the trend that giants emit more light in redder wavelengths than dwarfs. To ensure this trend is not an effect of extinction, we compared the average extinction of each template in each band (provided by the SDSS database). We find the difference in average extinction between the dwarf and giant templates to be smaller than the propagated errors in extinction provided by SDSS, leading us to conclude that any significant change in color is not an effect of extinction and thus, a real physical effect. For the higher temperature stars (F/G), we can only see the trend for bins of low metallicity because we do not have many templates for high-metallicity giants at high temperatures, and for low-temperature stars (K/M) we can only see the trend for high metallicities because we do not have many low-temperature, low-metallicity dwarf templates. However, with a complete set of data, we expect this trend to persist for all metallicity bins. We see the most separation between dwarf and giant stars for the lowest temperature stars (K/M). K and M stars contain many large molecular features and absorption lines, so this prominent color difference can be attributed to the surface gravity affecting the strength of many of these lines. The higher temperature stars do not show any separation in most cases (A-type, G/K-type), and a slight separation around $i-z = 0.1$ (F-type stars). 

\begin{deluxetable*}{c c c c c c c c c c c c c c} 
\tablecolumns{14}
\centering
\tablewidth{0pt}
\tablecaption{Photometry of Templates}
\tablehead{\colhead{Spectral Type} & \colhead{[Fe/H]} & \colhead{\textit{u-g}} & \colhead{$RMS_{u-g}$} &  \colhead{$\sigma_{u-g}$} &  \colhead{\textit{g-r}} & \colhead{$RMS_{g-r}$} &  \colhead{$\sigma_{g-r}$} & \colhead{\textit{r-i}} & \colhead{$RMS_{r-i}$} &  \colhead{$\sigma_{r-i}$} &  \colhead{\textit{i-z}} & \colhead{$RMS_{i-z}$} &  \colhead{$\sigma_{i-z}$}}           

\startdata   
             
O5V &-& -0.24& 0.14& 0.02& -0.51& 0.02& 0.02& -0.36& 0.02& 0.02& -0.31& 0.02& 0.02\\
O7V &-& -0.31& 0.05& 0.01& -0.47& 0.01& 0.01& -0.32& 0.01& 0.01& -0.34& 0.02& 0.02\\
A3V & +1.0& 1.2& 0.04& 0.02& -0.02& 0.05& 0.02& -0.08& 0.01& 0.02& -0.05& 0.04& 0.03\\
A3V & 0.0& 1.17& 0.01& 0.01& -0.07& 0.03& 0.01& -0.12& 0.02& 0.01& -0.07& 0.02& 0.01\\
A3V & -0.5& 1.12& 0.04& 0.02& -0.09& 0.02& 0.01& -0.1& 0.02& 0.01& -0.09& 0.02& 0.02\\
M0V & 1.0& 2.84& 0.05& 0.04& 1.29& 0.05& 0.01& 0.57& 0.03& 0.01& 0.3& 0.02& 0.01\\
M0V & 0.5& 2.59& 0.03& 0.01& 1.28& 0.01& 0.0& 0.58& 0.01& 0.0& 0.35& 0.01& 0.0\\
M0V & 0.0& 2.46& 0.03& 0.01& 1.26& 0.01& 0.0& 0.56& 0.01& 0.0& 0.33& 0.01& 0.0\\
F0III & -0.5& 1.25& 0.01& 0.02& 0.27& 0.0& 0.01& 0.07& 0.02& 0.01& 0.05& 0.0& 0.02\\
F0III & -1.0& 1.21& 0.03& 0.01& 0.2& 0.03& 0.01& 0.05& 0.02& 0.01& 0.01& 0.01& 0.01\\
F0III & -1.5& 1.16& 0.05& 0.01& 0.18& 0.03& 0.01& 0.04& 0.02& 0.01& 0.02& 0.01& 0.01\\ 
F1III & -1.0& 1.29& 0.0& 0.02& 0.13& 0.0& 0.01& -0.04& 0.0& 0.01& 0.01& 0.0& 0.02\\
\enddata 
\label{table:photometry} 
\end{deluxetable*}

\section{Conclusions}
\label{conclusions}

Using data taken with the SDSS BOSS spectrograph, we have compiled a new empirical stellar template library. 
Our template library: 
\begin{itemize}
\item Covers spectral types O5 through L3
\item Includes dwarf and giant separation for spectral types A0 through M8
\item Contains metallicity [Fe/H] bins for spectral types A3 through M8
\item Reports averaged photometric colors (in Sloan bands) for all the co-added stars in each template, along with a propagated errors and standard deviations
\end{itemize} 

Along with the templates, we have released the PyHammer code for assigning a spectral type and metallicity automatically (or by visual inspection). This code is based on the ``Hammer" spectral typing facility, written by \citet{covey07}, but includes metallicity information and is now written in Python. The automatic spectral typing portion of code returns the exact spectral type we determined using the original ``Hammer" code and metallicity we determined using the methods described in Section \ref{metallicity} over 50\% of the time. The spread in the spectral type was 1.5 spectral sub-types, and a spread in the metallicity was 0.4 dex. Visual spectral typing allows for direct comparison between input spectra and our empirical templates in an easy to use GUI. The code is available on GitHub\footnote[4]{github.com/BU-hammerTeam/PyHammer}. 

The library of empirical stellar spectra will be important for a wide range of research topics from extragalactic to galactic astronomy, planetary system stellar characterization, and even as an astronomical teaching tool. With large photometric surveys such as LSST, machine learning techniques will become increasingly important to quickly characterize large amounts of data. Along with releasing our templates, we will provide lists of the individual BOSS spectra co-added to construct each template. This combination of information will be an ideal training set for machine learning, and can extend the work of \citet{miller15} on F, G, and K stars to both higher and lower mass stars. The templates also provide the necessary tool for characterizing stellar populations in other galaxies, especially for studies of the IMF. Scrutiny of the low-mass end of the IMF, has led many people to suggest it changes form in different environments (i.e. different metallicity environments). Our catalog represents the first empirical template library with metallicity and surface gravity separation for low-mass (M-type) stars. The catalog and the new ``PyHammer" spectral typing facility will be a useful tool for the community as a whole. 

\acknowledgements
The authors would like to thank Phil Muirhead and Chris Theissen for reading this manuscript and providing much helpful feedback. The authors would also like to thank Dylan Morgan, Chris Theissen, Conor Robinson and Philip Phipps for help developing the ``PyHammer" code. 
A.A.W. also acknowledges the support of the NSF grants AST-1109273 and AST-1255568 along with Research
Corporation for Science Advancement's Cottrell Scholarship.

Funding for the SDSS and SDSS-II has been provided by
the Alfred P. Sloan Foundation, the Participating Institutions,
the National Science Foundation, the U.S. Department of Energy,
the National Aeronautics and Space Administration, the
Japanese Monbukagakusho, the Max Planck Society, and the
Higher Education Funding Council for England. The SDSS Web
Site is $http://www.sdss.org/$. The SDSS is managed by the Astrophysical
Research Consortium for the Participating Institutions.
The Participating Institutions are the American Museum
of Natural History, Astrophysical Institute Potsdam, University
of Basel, University of Cambridge, Case Western Reserve
University, University of Chicago, Drexel University, Fermilab,
the Institute for Advanced Study, the Japan Participation
Group, Johns Hopkins University, the Joint Institute for Nuclear
Astrophysics, the Kavli Institute for Particle Astrophysics and
Cosmology, the Korean Scientist Group, the Chinese Academy
of Sciences (LAMOST), Los Alamos National Laboratory, the
Max Planck Institute for Astronomy (MPIA), the Max Planck
Institute for Astrophysics (MPA), New Mexico State University,
Ohio State University, University of Pittsburgh, University
of Portsmouth, Princeton University, the United States Naval
Observatory, and the University of Washington.

This publication makes use of data products from the Two
Micron All-Sky Survey, which is a joint project of the University
of Massachusetts and the Infrared Processing and Analysis
Center/California Institute of Technology, funded by the
National Aeronautics and Space Administration and the National
Science Foundation. This publication also makes use of
data products from the Wide-field Infrared Survey Explorer,
which is a joint project of the University of California, Los
Angeles, and the Jet Propulsion Laboratory/California Institute
of Technology, funded by the National Aeronautics and Space
Administration.

\section{Appendix A} 
In this appendix, we explain in detail the automatic, and visual spectral typing code, dubbed ``PyHammer"  \citep[based on the ``Hammer" by ][]{covey07}.  The code automatically outputs best estimates for the radial velocity, spectral type, and metallicity (when metallicity information is available: A3 - M8). We also present the methods for determining these estimates, tests to determine the accuracy/precision of these methods, and the resulting accuracy for each parameter determined. There is also a visual, or ``by-eye" spectral typing feature, which allows the user to visually compare the input spectrum to any of the template spectra in a GUI window. The PyHammer code is available on GitHub\footnote[5]{github.com/BU-hammerTeam/PyHammer} and we have set up an email address for questions, comments and suggestions\footnote[6]{astro.pyhammer@gmail.com}. 

The general procedure of the code is to first interpolate the input spectrum onto the same wavelength grid as our templates and convert to vacuum wavelengths (if necessary) to allow for direct comparison. We then measure 34 spectral indices and color regions given in Table \ref{table:hammer} (all in vacuum wavelengths) and make an initial estimate of the spectral type using these indices. With that spectral type estimate, we can determine the radial velocity by cross correlating the spectrum with the corresponding template. We then shift the input spectrum to its rest frame, and re-measure the 34 spectral indices to determine a more accurate estimate of the spectral type and metallicity. 

\subsection{Radial Velocity} 

The radial velocity cross correlation method is based on an IDL procedure \texttt{xcorl} \citep{mohanty03, west09}, which was translated into python by \citet{theissen14}. The cross correlation method examines three regions of the spectrum (5000-6000 \AA, 6000-7000 \AA, and 7000-8000 \AA), and for each region performs a cross correlation. The shift that produces a minimum in the cross-correlation function is recorded, and a sigma-clipped median of the measurements from all the regions is reported as the radial velocity. 

\begin{figure*}[]
\begin{center}
\includegraphics[scale=0.4]{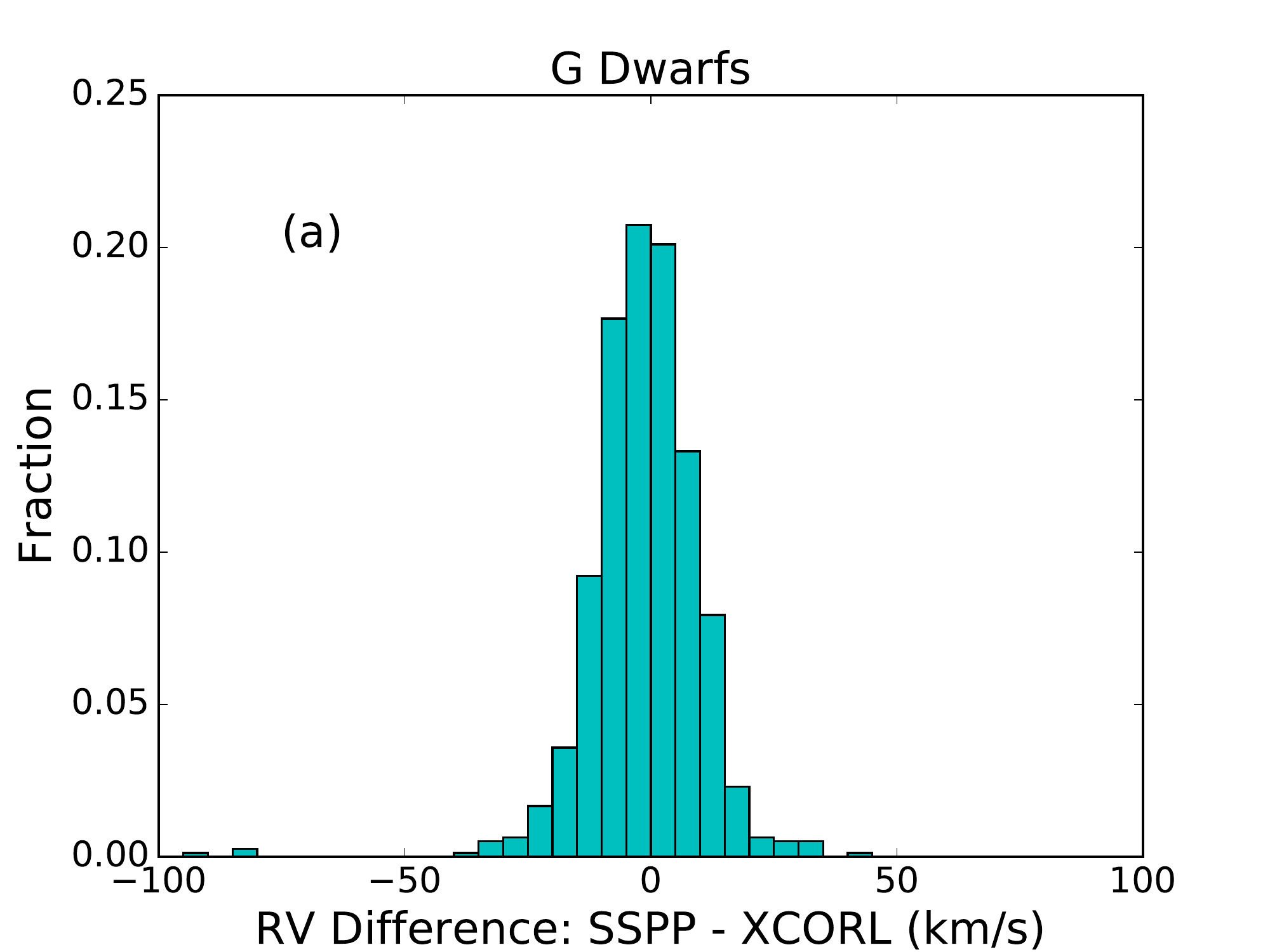}
\includegraphics[scale=0.4]{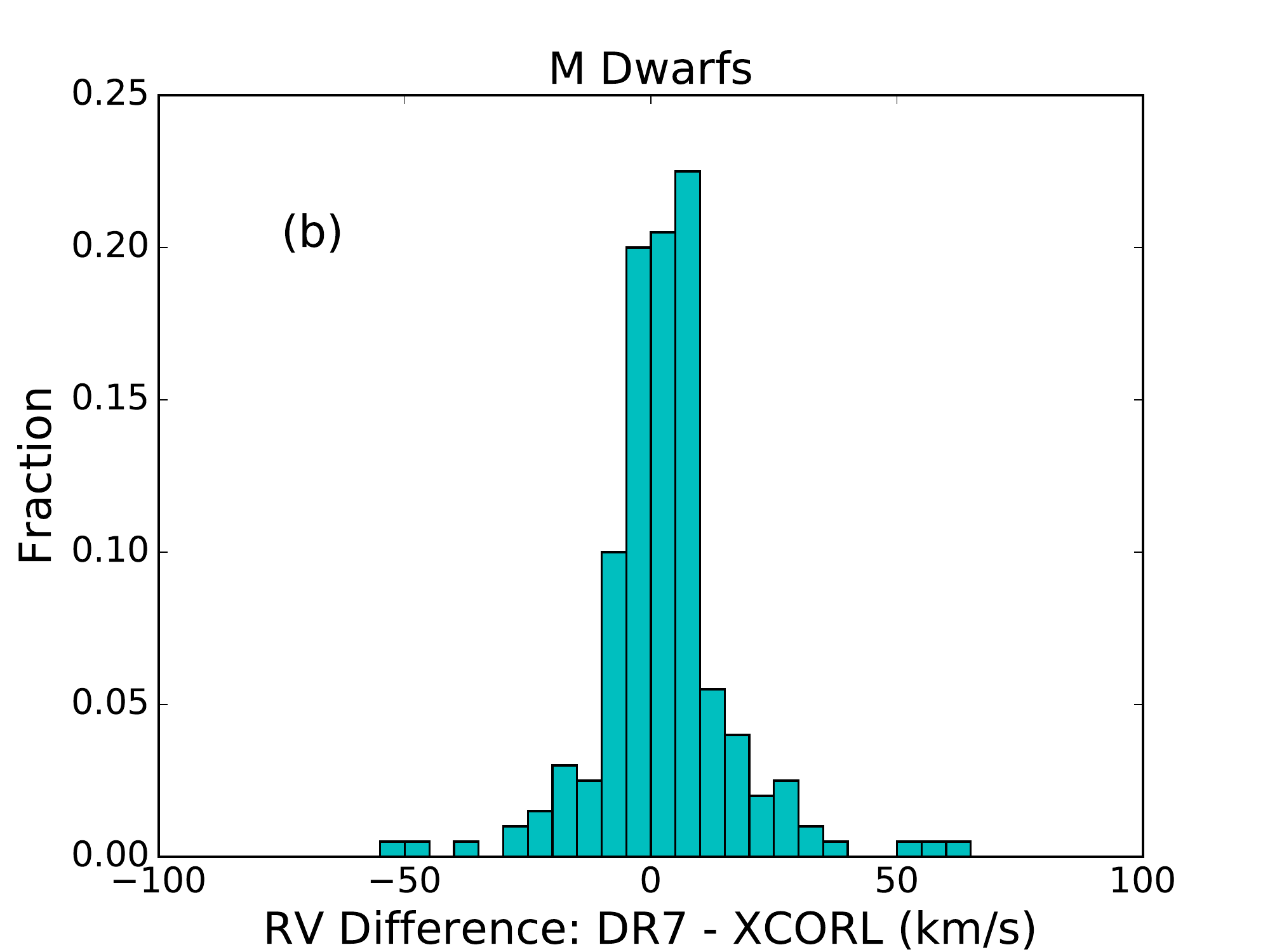}
\includegraphics[scale=0.4]{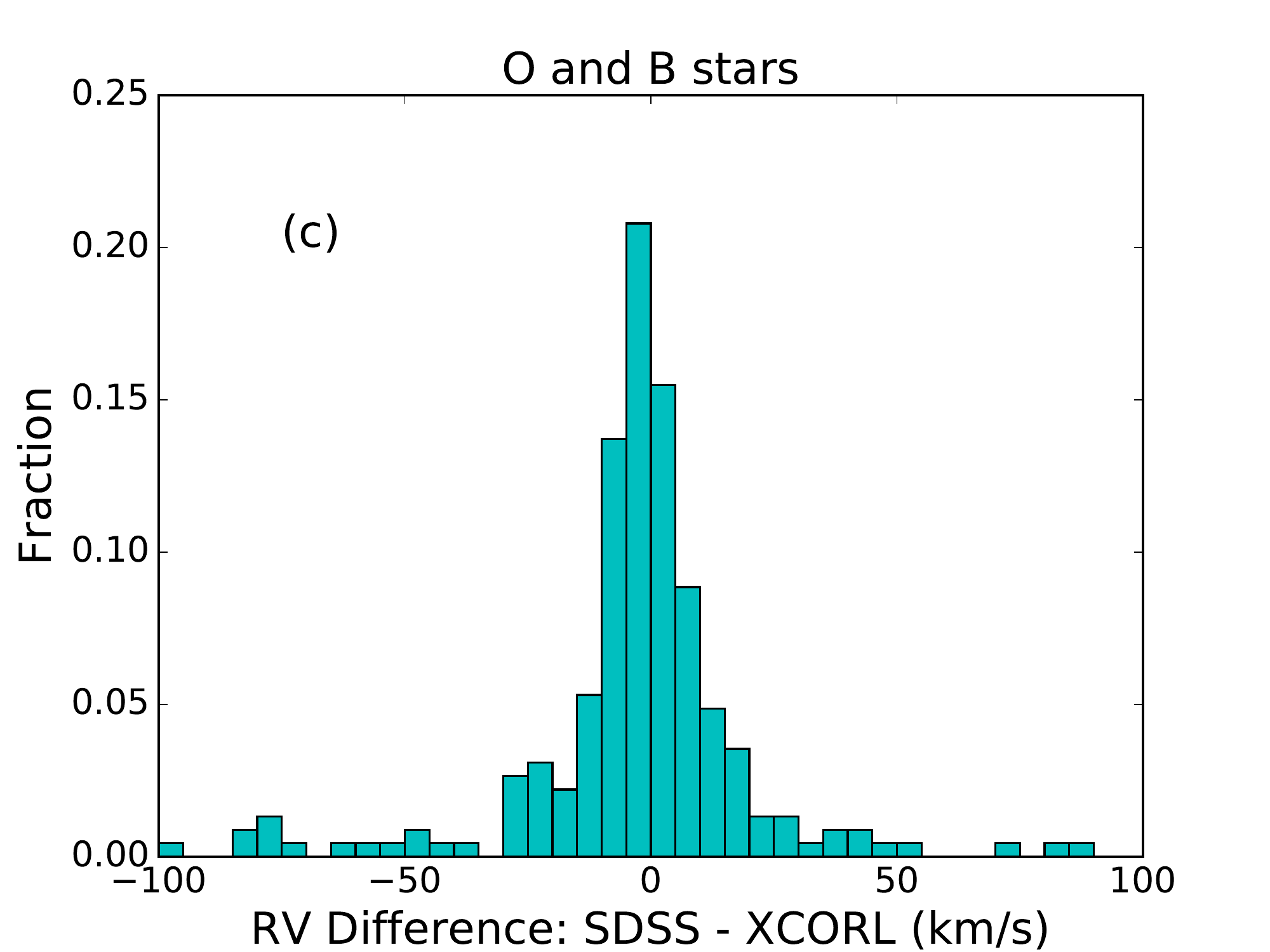}
\caption{\small 
Comparison between our measured radial velocities and \textbf{(a)} the radial velocities determined by the SSPP for the G dwarfs, \textbf{(b)} by cross correlation with \citet{bochanski07} templates using \citet{west11} DR7 sample for the M dwarfs, and \textbf{(c)} the SDSS pipeline for the O and B stars. The radial velocities for the G stars have a median difference of -1 km s$^{-1}$ and ``1-sigma" (68\%) of the data falls between -10.0 and 7.9 km s$^{-1}$. The M dwarfs have a median difference of 2 km s$^{-1}$ and 68\% of the data fall between -7.1 and 10.8 km s$^{-1}$. The O/B stars are significantly more spread out, and have a median difference of -1.6 km s$^{-1}$, with upper and lower limits of ``1-sigma" at -22.7 and 10.7 km s$^{-1}$.
Our radial velocities measurements are therefore comparable in uncertainty to the previous methods for calculating radial velocities.
}
\label{f:rv}
\end{center}
\end{figure*}

To test the radial velocity code, we compared our measured radial velocities to previously measured radial velocities across the full spectral range. Since there were three different methods (one for O, B, and A-type stars, one for F, G, K-type stars, and one for M-type stars) used for the initial radial velocity calculations, each with different uncertainties, we chose a sample of spectra in each of the temperature regimes with which to compare our radial velocities. For the F, G and K-type stars, we compared our radial velocities to radial velocities derived in the SSPP, which are accurate to $\sim$7--9 km s$^{-1}$ (see Section \ref{FGK:RV}). To validate the radial velocities for the low-temperature stars, we compared the radial velocities of the \citet{west11} sample of M dwarfs, derived by cross correlation with \citet{bochanski07} M dwarf templates, to our measured radial velocities. The M dwarf radial velocities reported in the sample (BOORV field) are accurate to $\sim$10 km s$^{-1}$. For the higher temperature stars, we compared our radial velocities to the radial velocities derived from the SDSS pipeline, which are accurate to 10 --15 km s$^{-1}$ (see Section \ref{OBA:RV}), making this comparison the least accurate. The results are shown in Figure \ref{f:rv}. 

For all the comparisons the difference between the two radial velocities peaks around 0 km s$^{-1}$. 
Instead of computing a simple standard deviation, we recorded the value which encloses 34\% of the data both above and below the median. We find this to be more representative than a standard deviation because the distributions are not true gaussians (especially for the O and B-type stars). 
We find a median value of -1.2 km s$^{-1}$ between our measurements and the SSPP values for the G dwarfs. The ``1-sigma" spreads above and below the median are 7.9 km s$^{-1}$ and -10 km s$^{-1}$ respectively. For the low-temperature stars, we find a median of 2.2 km s$^{-1}$, a ``1-sigma" upper and lower bound of 10.8 km s$^{-1}$ and -7.1 km s$^{-1}$ respectively. The high-temperature stars have the lowest precision,  with a median of -1.6 km s$^{-1}$, an upper bound of 10.7 km s$^{-1}$ and a lower bound of -22.7 km s$^{-1}$. The high-temperature distribution contains many outliers, which we believe is due to a combination of higher uncertainty in the SDSS pipeline for high-mass stars than the 10 --15 km s$^{-1}$ quoted values, and higher errors in our calculated radial velocities because of fewer absorption lines (especially in O stars). 
Our analysis of Figure \ref{f:rv} leads us to conclude that our radial velocity measurements have comparable uncertainties to those given by previous measurements for mid-temperature and low-temperature stars, with an uncertainty of $\sim 7-10$ km s$^{-1}$. The radial velocities of the high-temperature stars are on par or slightly less accurate than previous measurements, with uncertainties on the order of 15-20 km s$^{-1}$.

\subsection{Spectral Type/Metallicity Estimate} 
\label{pyhammerEstimates}

To assign an initial estimate of the spectral type and metallicity, we compared the input spectrum to the template spectra in the following manner. We first measured 34 spectral indices (Table \ref{table:hammer}) from each individual spectrum that was used to create the templates. For all spectra of a given spectral type and metallicity (i.e., each template), we measured the weighted mean and variance of each index for all of the co-added spectra, where the weight is the variance in the index value due to uncertainties in the observed spectrum. We repeated this process for all spectral types and metallicities. To reduce computation time, the results of this procedure were saved to an external file and not repeated each time the code was run.

We compared the input spectrum to each template by computing a chi-squared value that compares the spectral indices measured from the input spectrum to the weighted mean indices described above. The variance used in the chi-squared calculation is the variance among indices of multiple stars of the same template type as described above. We reported the spectral type and metallicity of the template that produces the minimum chi-squared as the initial estimate. We chose to use indices instead of doing a chi-squared minimization of the entire spectrum because spectral indices are normalization independent, robust against missing data, and take less computational time. 



\begin{figure*}[]
\begin{center}
\includegraphics[scale=0.4]{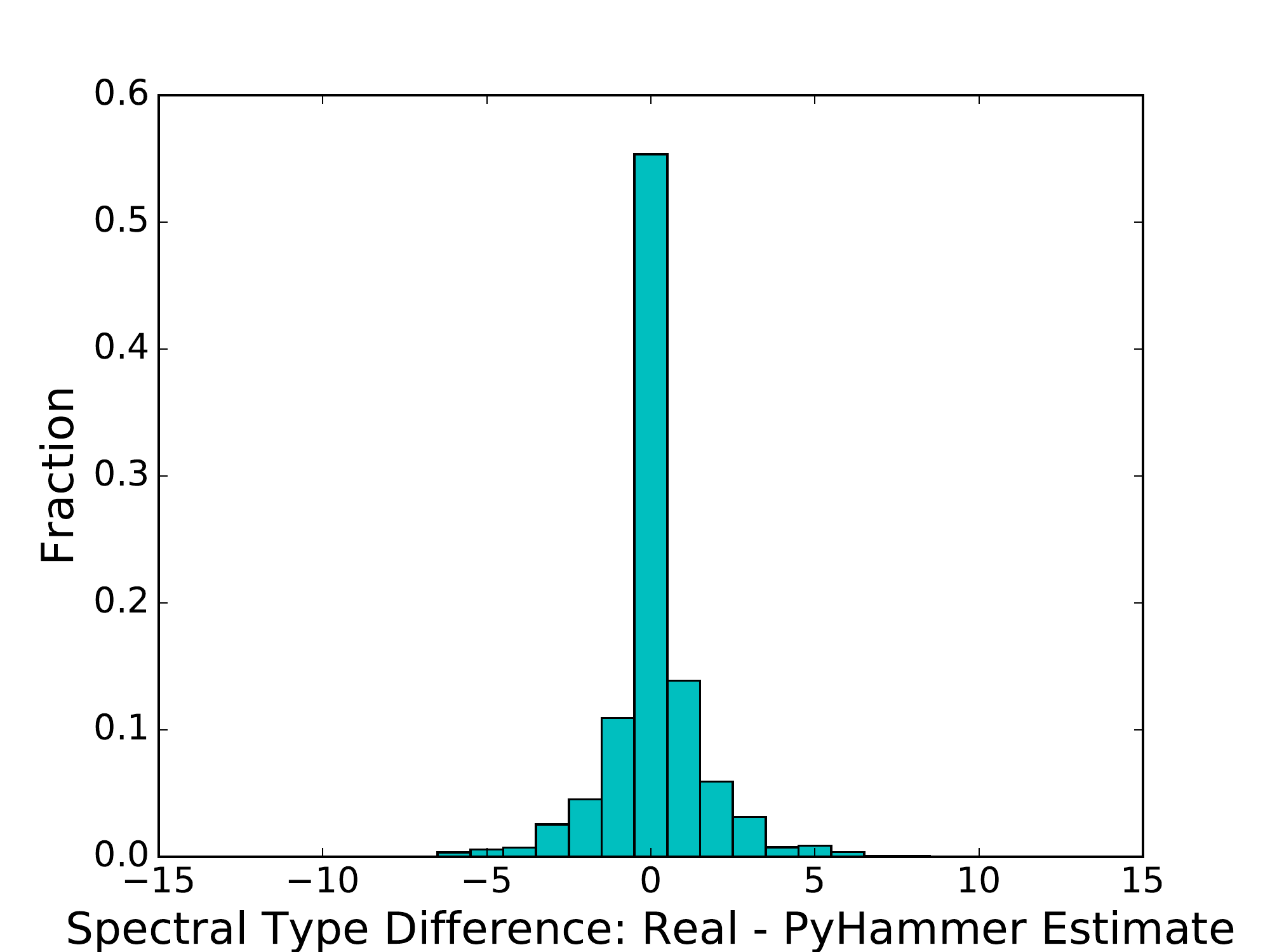}
\includegraphics[scale=0.4]{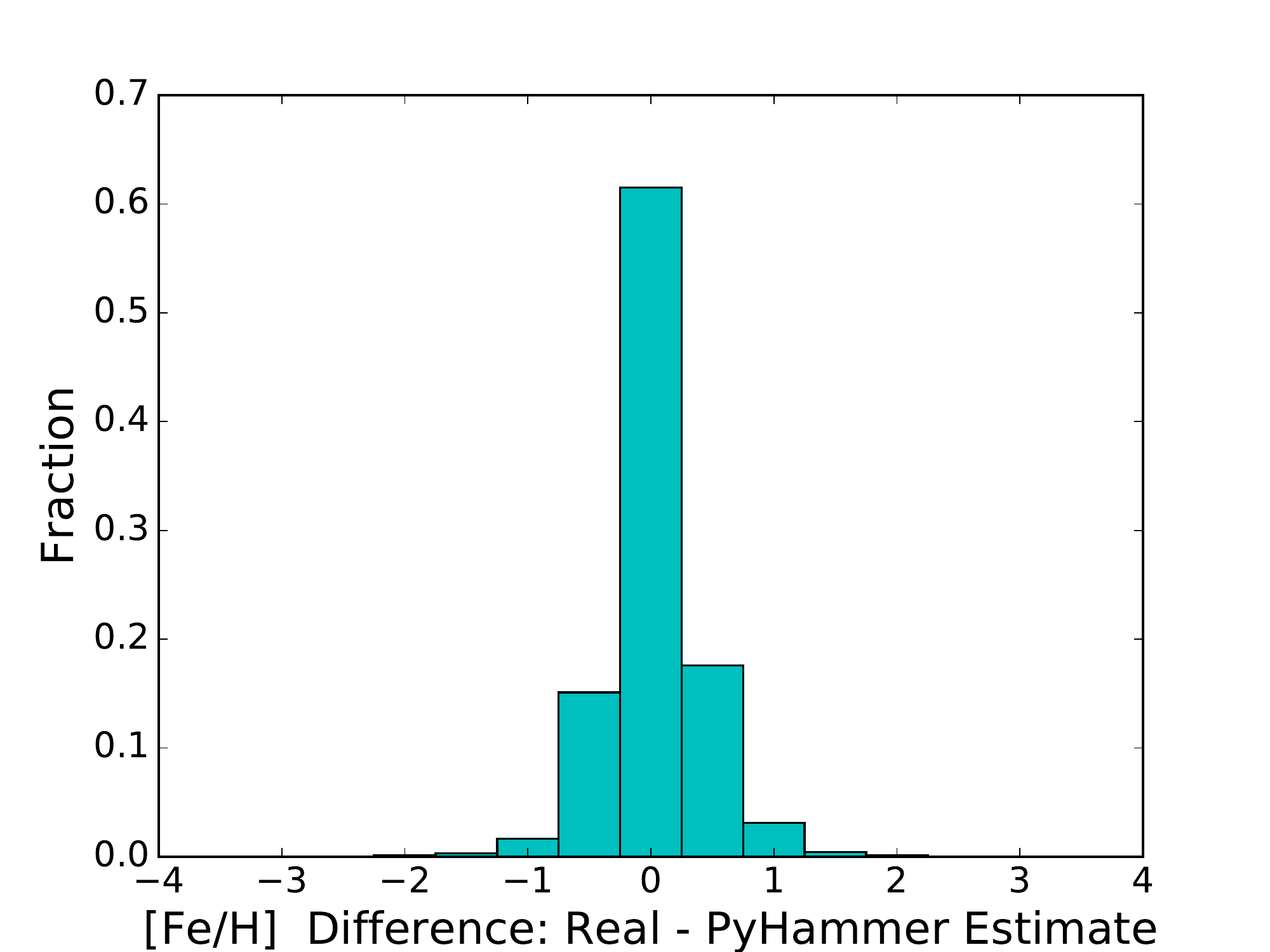}
\caption{\small 
Comparison between the hammer automatically estimated spectral types (left) and metallicities (right) and the actual spectral types and metallicities we determined for the individual BOSS spectra that were used to make the templates.  The spectral type estimate returns the correct value more than 50\% of the time, and the metallicity estimate returns the correct value upwards of 60\% of the time. The spread in the spectral type is extremely small, with a standard deviation of 1.5. The standard deviation in the metallicity is 0.4 dex, which is smaller than one of our metallicity bins. 
}
\label{f:spt_compare}
\end{center}
\end{figure*}

We tested the automatic spectral typing by running the code on the individual spectra that were used in constructing the templates and compared the estimated metallicity and spectral type to the actual metallicity and spectral type we determined. The results are shown in Figure \ref{f:spt_compare}. Both comparisons show good agreement, with over 50\% of the estimates exactly the same as the determined spectral types and metallicities. The spread in both comparisons is also extremely small, with the standard deviation being 1.5 spectral subtypes in the spectral-type comparison, and 0.4 dex in the metallicity comparison. We conclude that even without the additional visual inspection, the automatic spectral typing and metallicity estimates are within one metallicity bin and one spectral subtype over 80\% of the time. 

\begin{deluxetable*}{c c c l}
\scriptsize
\tablecolumns{3}
\centering
\tablewidth{0pt}
\tablecaption{Spectral Indices Used by PyHammer} 
\tablehead{\colhead{Spectral Index} & \colhead{Numerator (\AA)} & \colhead{Denominator (\AA)}}
\startdata
Ca II K & 3923.7 - 3943.7 & 3943.7 - 3953.7\\
H$\delta$ & 4086.7 - 4116.7& 4136.7 - 4176.7 \\
Ca I 4227 & 4216.7 - 4236.7& 4236.7 - 4256.7 \\
G-band & 4285.0 - 4315.0& 4260.0 - 4285.0 \\
H$\gamma$ & 4332.5 - 4347.5& 4355.0 - 4370.0 \\
Fe I 4383 & 4378.6 - 4388.6& 4355.0 - 4370.0 \\
Fe I 4404 & 4399.8 - 4409.8& 4414.8 - 4424.8 \\
H$\beta$ & 4847.0 - 4877.0& 4817.0 - 4847.0 \\
Mg I & 5152.7 - 5192.7& 5100.0 - 5150.0 \\
Na D & 5880.0 - 5905.0& 5910.0 - 5935.0 \\
Ca I 6162 & 6150.0 - 6175.0& 6120.0 - 6145.0 \\
H$\alpha$ & 6548.0 - 6578.0& 6583.0 - 6613.0 \\
CaH2 & 6814.0 - 6845.0 & 7042.0, 7046.0\\
CaH3 & 6960.0 - 6990.0& 7042.0 - 7046.0 \\
TiO5 & 7126.0 - 7135.0& 7042.0 - 7046.0 \\
VO 7434 & 7430.0 - 7470.0& 7550.0 - 7570.0 \\
VO 7445 & 7350.0 - 7400.0, 0.5625\tablenotemark{1}; 7510.0 - 7560.0, 0.4375 & 7420.0 - 7470.0 \\
VO-B & 7860.0 - 7880.0, 0.5; 8080.0 - 8100.0, 0.5& 7960.0 - 8000.0 \\
VO 7912 & 7900.0 - 7980.0& 8100.0 - 8150.0 \\
Rb-B & 7922.6 - 7932.6, 0.5; 7962.6 - 7972.6, 0.5 & 7942.6 - 7952.6 \\
Na I & 8177.0 - 8201.0& 8151.0 - 8175.0 \\
TiO8 & 8400.0 - 8415.0& 8455.0 - 8470.0 \\
TiO 8440 & 8440.0 - 8470.0& 8400.0 - 8420.0 \\
Cs-A & 8496.1 - 8506.1, 0.5; 8536.1 - 8546.1, 0.5 & 8516.1 - 8526.1 \\
Ca II 8498 & 8483.0 - 8513.0& 8513.0 - 8543.0 \\
CrH-A & 8580.0 - 8600.0& 8621.0 - 8641.0 \\
Ca II 8662 & 8650.0 - 8675.0& 8625.0 - 8650.0 \\
Fe I 8689 & 8684.0 - 8694.0& 8664.0 - 8674.0 \\
FeH &9880.0 - 10000.0 & 9820.0 - 9860.0 \\
Color Region 1 & 4550-4650 & 4160-4210 \\
Color Region 2 & 5700-5800 & 4160-4210 \\
Color Region 3 & 7480-7580 & 4160-4210 \\
Color Region 4 & 9100-9200 & 4160-4210 \\
Color Region 5 & 10100-10200 & 4160-4210\\ 
\enddata

\tablenotetext{1}{\footnotesize Indices with more than one numerator entry contain two numerator regions and a weight for each region, e.g. lower limit of region 1 -- upper limit region 1, weight of region 1; lower limit of region 2 -- upper limit of region 2, weight of region 2}
\label{table:hammer}
\end{deluxetable*}

\subsection{Running PyHammer}
\normalsize
Detailed instructions on running PyHammer as well as more advanced features of the code are provided in the README file in GitHub, and in the `Help' menu (available when the visual classification GUIs are displayed). Here, we will give a brief overview of how to run the code and a description of some of the features available. After starting the code (type ``python pyhammer.py" on the command line), an initial GUI window will appear, allowing the user to enter the name of (or create) the input and output files, specify a path to the spectra files, skip straight to the by-eye spectral typing (as opposed to doing the automatic spectral typing first), and apply a signal-to-noise cut-off. Skipping directly to the eye check stage should only be done if the automatic spectral typing has already been completed at an earlier time. If the user does not skip the automatic spectral typing, the program determines the radial velocity, metallicity and spectral type and writes these to the output file (PyHammerResults.csv by default). If the user supplies a signal-to-noise cut-off, any spectrum with a median signal to noise ratio (calculated simply as the median flux divided by the median uncertainty over the entire spectrum) lower than the specified value is written to RejectSpectra.csv. There is an example input file (exampleInputFile.txt) and a few spectra in the test\_case folder to demonstrate how the input file should be set up, and to aid the first time user. 

\begin{figure*}[]
\begin{center}
\includegraphics[scale=0.5]{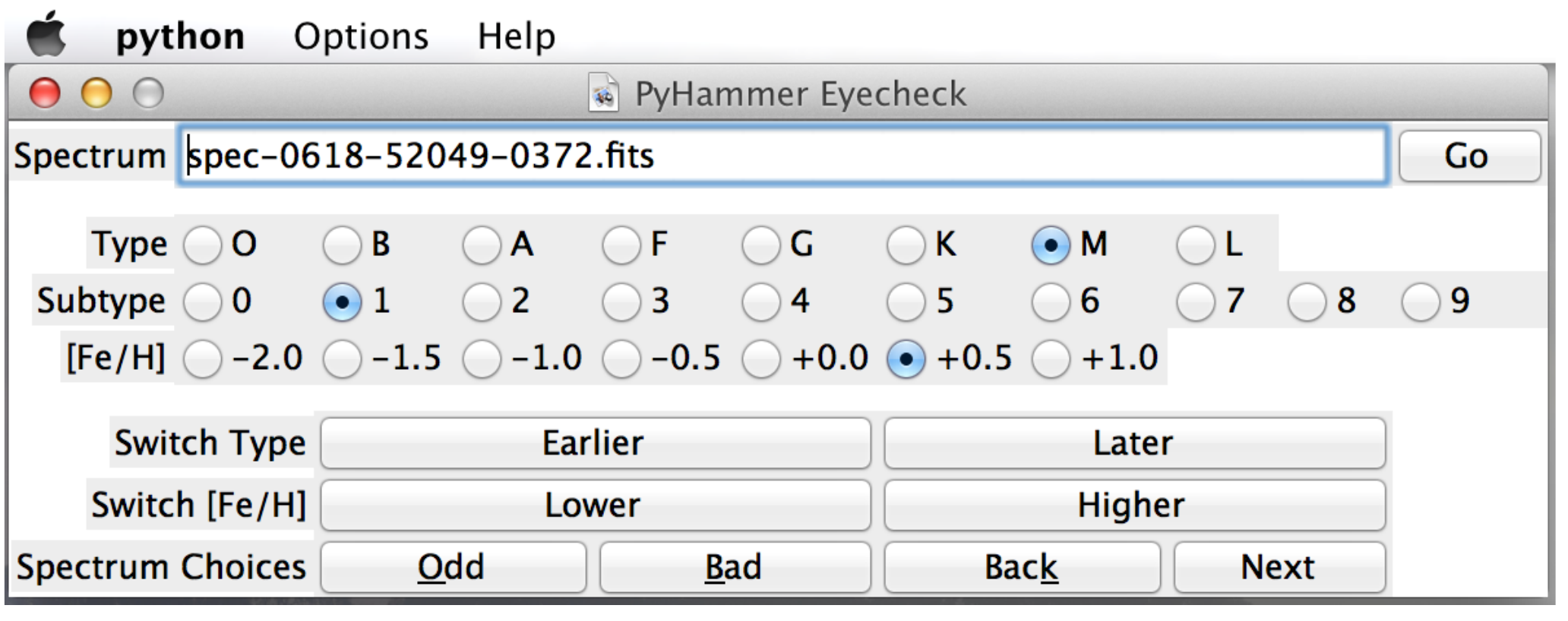}
\includegraphics[scale=0.5]{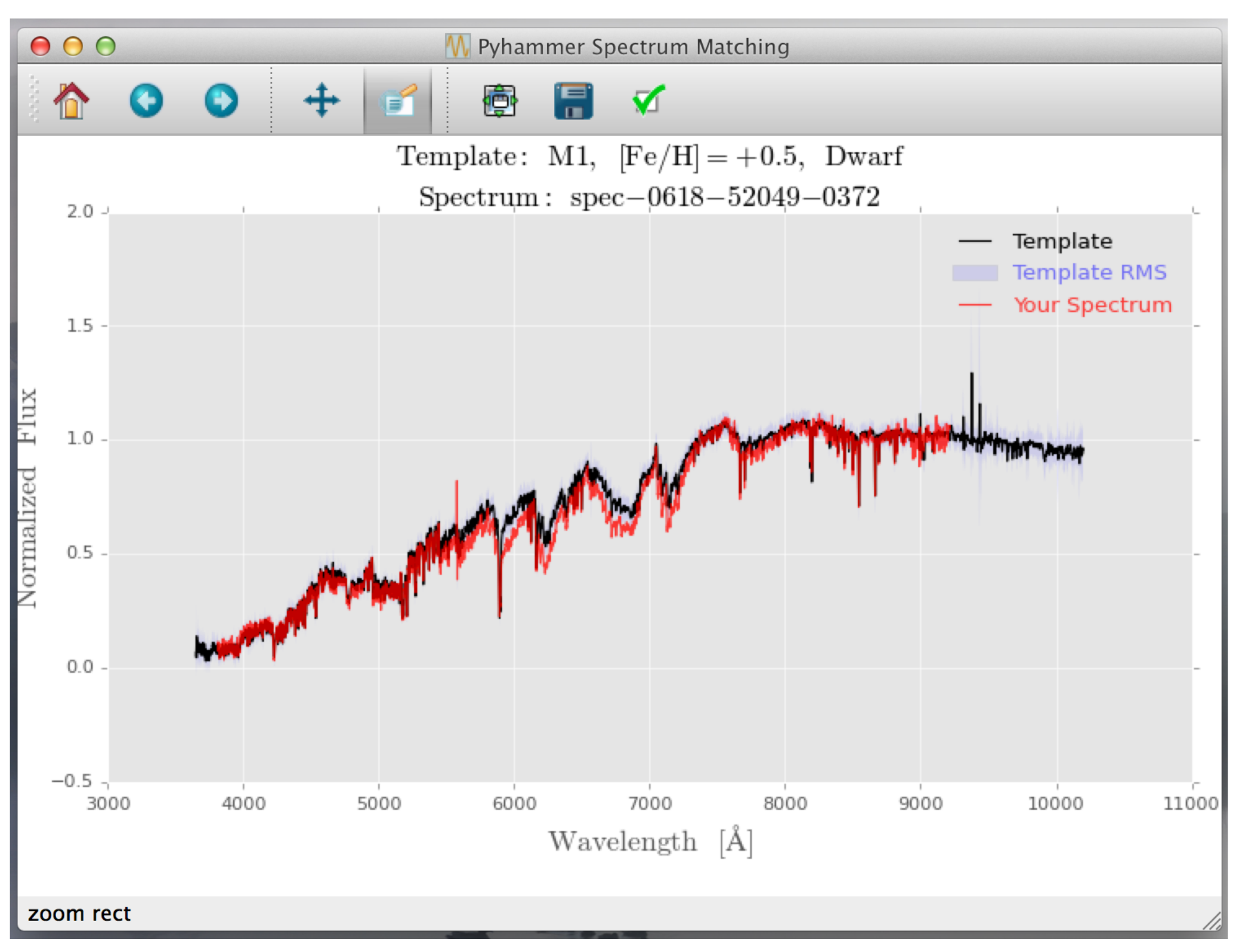}
\caption{\small 
Screenshot showing the two GUIs used for the by-eye spectral typing. The top GUI screen allows the user to display templates of different spectral type and metallicity. The bottom GUI screen initially displays the best guess template (in black), along with the standard deviation from all the co-added individual spectra at each wavelength (in semi-transparent blue). The spectrum from the user's input list is displayed in red.
}
\label{f:GUI}
\end{center}
\end{figure*}

A screenshot of the GUI for the visual spectral typing is shown in Figure \ref{f:GUI}. The GUI allows the user to display the spectra and the templates on the same plot, allowing for direct comparisons. All of the normal matplotlib graphing buttons are functional, and shown in the top left corner of the GUI in Figure \ref{f:GUI}. By pushing the magnifying glass button, the user can zoom in on specific regions of the spectrum. While zoomed in the user can use the four sided arrow to scroll. Finally, by pressing the home button, the graph will be taken to the original view. In the `Options' menu at the top of the screen there are four different viewing options for the graph. The user can choose to display or not display the template RMS (in transparent blue). The user can also smooth the spectrum, and lock the smoothing since by default every new spectrum loaded will be unsmoothed. The smoothing option runs a simple boxcar smoothing over the input spectrum, which reduces noise and allows for easier comparisons for low signal to noise spectra. Finally if the spectra are from SDSS, the user can choose to remove the known false spike in the spectra at 5580 \AA, created by the stitching together of the red end and blue end of the spectra.

Along with simply clicking through different templates on the bottom part of the GUI shown in Figure \ref{f:GUI}, there are many options designed to make visual spectral typing easier. The `Earlier' and `Later' buttons change the spectral subtype by one each time, where `Earlier' scrolls to lower numbered subtypes (e.g., M4 to M3 or M0 to K7), and the `Later' to higher numbered subtypes. While the `Lower' and `Higher' buttons change the metallicity by one bin (0.5 dex) with each click. There are also buttons with other options, which have been adopted from the original Hammer code. The `Odd' button allows the user to input their own note into the space where the spectral type would normally be stored. We have set a few standard `odd' spectral types like white dwarf (Wd), white dwarf+M dwarf binary (Wdm), carbon star, galaxy (Gal) and unknown, but the user may also type anything he or she would like into this space. Finally, the user can proceed to the next spectrum from their input list or return to a previous spectrum using the `Back', `Next' buttons. Each of these buttons also has a keyboard shortcut so the entire visual spectral typing can be done with only a keyboard (to speed up the process). For more information on the keyboard shortcuts go to the `Help' menu and click on the `Keys' tab in the PyHammer Help window. 

\end{document}